\documentclass{jfm}
\usepackage{graphicx}
\usepackage{epstopdf, epsfig}
\usepackage{amsmath}
\newcommand{\vx}{{\mathbf x}}

\newcommand{\va}{{\mathbf a}}

\shorttitle{Transport under advective trapping}
\shortauthor{J. J. Hidalgo, I. Neuweiler and M. Dentz}

\title{Transport under advective trapping}

\author{Juan J. Hidalgo\aff{1}
  \corresp{\email{juanj.hidalgo@idaea.csic.es}},
  I. Neuweiler\aff{2}
 \and M. Dentz\aff{1}}

\affiliation{\aff{1} IDAEA-CSIC, Barcelona, 08034, Spain
  \aff{2} Leibniz Universit{\"a}t Hannover, Hannover, Germany}

\begin{document}
\maketitle
\begin{abstract}
  Advective trapping occurs when solute enters low velocity zones in
  heterogeneous porous media. Classical local modeling approaches
  combine the impact of slow advection and diffusion into a
  hydrodynamic dispersion coefficient and many temporally non-local
  approaches lump these mechanisms into a single memory function. This
  joint treatment makes parameterization difficult and thus prediction
  of large scale transport a challenge. Here we investigate the
  mechanisms of advective trapping and their impact on transport in
  media composed of a high conductivity background and isolated low
  permeability inclusions. Breakthrough curves show that effective
  transport changes from a streamtube-like behavior to genuine random
  trapping as the degree of disorder of the inclusion arrangement
  increases. We upscale this behavior using a Lagrangian view point,
  in which idealized solute particles transition over a fixed distance
  at random advection times combined with Poissonian advective
  trapping events. We discuss the mathematical formulation of the
  upscaled model in the continuous time random walk and mobile-immobile
  mass transfer frameworks, and derive a model for large scale solute
  non-Fickian dispersion. These findings give new insight into
  transport in highly heterogeneous media.

\end{abstract}
%
%
\section{Introduction}
\label{sec:intro}
Predicting flow and transport processes in the subsurface is
challenging, as the heterogeneous subsurface structure is usually not known. 
Heterogeneity can cause 
a broad distribution of transport time scales with short times for advective transport
along fast paths and very long times for diffusive and
advective transport in the zones with very low to zero flow velocity
\citep{BS1997,Jardineetal99, Haggerty2000}. Depending on the
subsurface structure, the full range of time scales can be important
for scalar transport. Although the larger fraction of the mass
might be transported fast, a substantial fraction can experience very
large transport times, which might be crucial for applications such as
contaminant remediation or recovery of substances. The range of
transport time scales causes challenges for predictions. If the
subsurface structure is known, numerical solutions of the transport
equation in the domain can be derived. The computational effort is,
however, very high, as the resolution of all relevant time and spatial
scales is required. If the structure is not known, statistical
approaches might be used, which increases the computational burden
even more.

Upscaled transport models are derived in order to allow for efficient
predictions, where the detailed resolution is not required, but the
effects of the non-resolved processes are captured in effective
transport mechanisms. The derivation of upscaled transport equations
has been pursued in the frameworks of volume averaging~\citep{brenner:book,Whitaker:book},
homogenization theory~\citep{Hornung:book}, and stochastic
averaging~\citep{NEUMAN1993,CBH2002}, which can yield local or spatio-temporal non-local upscaled
transport equations that typically rely on closure
approximations. Such closure approximations often rely on the
assumption of weak heterogeneity, or on the assumption that average
transport is Fickian. 

Mobile-immobile mass transfer (MIM), matrix-diffusion and multi-rate mass transfer (MRMT)
approaches derived for solute transport in highly heterogeneous media
conceptualize the medium as primary continuum and a suite of multiple
secondary continua
\citep{Haggerty1995, Carrera1998}. The fastest domain covers the main
transport, while the mass exchange with the other continua is
described as a source term. The source term is formulated as a
convolution of the concentration in the fast domain and a memory function. 
The memory function encodes the mass transfer processes between mobile
and immobile domains. An overview of the terminology of
mobile-immobile, multirate mass transfer and in general memory
function models can be found in~\cite{Ginn2017}. 
  
The continuous time random walk (CTRW) approach for
transport in highly heterogeneous media naturally accounts for broad
distributions of transport time scales over characteristic length
scales inherent to the medium structure~\citep{Berkowitz2006}. The
information about small scale mass transfer and medium
structure is contained in the transition time distribution. The
phenomenology of mobile-immobile and CTRW approaches is similar in that both
account for broad distributions of mass transfer time scales. In fact,
the mathematical equivalence between the frameworks has been shown in
the literature~\citep{Dentz2003, Schumeretal2003, BensonMeer2009,
 Russian2016, Comolli2016}.

A crucial point for an upscaled model is predictability. A model is
useful for applications if parameters can be identified
independently from specific settings. They should either be
predictable from knowledge about material properties and specific
transport parameters, or should be transferable, meaning that if they
are fitted from experimental observations, they should be transferable
to other settings. Comparative studies of the predictive capabilities of different
upscaling approaches and large scale models can be found in
\cite{Frippiat2008},~\cite{Neuman2008}, \cite{Fiori2015}, \cite{Lu2018}
and~\cite{Pedretti2018}. 

The parameterization of mobile-immobile models for the case that
transport in the slow domains is dominantly
diffusive has been studied in the past. There is a good
understanding of the memory function and how parameters can be
estimated based on diffusion coefficients and the geometry of
the heterogeneous medium (or fractured medium)
\citep{maloszewski1985,Carrera1998,Zhang2014,GMDC2008}. Oftentimes,
both slow advection and diffusion are lumped into empirical memory
functions based on parametric models like truncated power laws~\citep{Willmann2008}. 
It is not clear, however, whether slow advection can be represented in
such a framework, and what the form and parameterization of the memory
function would be. 

In general, both advective transport as well as diffusive
transport are relevant for the scalar transport in the slow zones
of a heterogeneous medium. To formulate mobile-immobile mass transfer
models in general requires a method to parameterize the memory
functions for a combination of diffusive and advective transport. As
mentioned above, in the MRMT framework the effects of advection and
diffusion have been accounted for by phenomenological memory
functions~\citep{Willmann2008}, and similarly, in the CTRW approach, the
combined effect of diffusion and advection on solute travel times have
been quantified single parametric transition time
distributions~\citep{Berkowitz2000}. Volume averaging has been used as
a systematic way to quantify transport and advective-diffusive mass
transfer in bimodal media~\citep{Chastanet2008, Golfier2011,
  Davit2012}, which, however, typically leads to more or less complex
closure problems. Closure approximations can be based on weak
heterogeneity~\citep{Golfier2011}, or the introduction of
time-dependent mass transfer coefficients~\citep{Chastanet2008}.

The impact of advective mass transfer between slow and fast medium
portions, can be systematically assessed by studying purely advective
transport in highly heterogeneous media. Thus, in order to 
investigate the mechanisms of advective trapping in heterogeneous
porous media, we focus here on structures characterized by a
background-inclusion pattern. The simplest model for such a structure
is a 2D medium with circular inclusions.

\citet{Eames1999} studied the advective transport
in a background-inclusion field with a bimodal conductivity
distribution. These authors consider regular,
as well as random structures of the inclusions. It is demonstrated in
their paper that the macrodispersion coefficient that is derived for
transport in such media diverges for the case that the inclusions are
permeable in the limit of an inclusion/matrix permeability ratio to
zero. If on the other hand the transport coefficient is calculated for
the case that inclusions are impermeable, a finite macrodispersion
coefficient is obtained. This observation indicates that the concept
of hydrodynamic dispersion is not adequate to describe transport in
the case of very low permeability ratios. 

\cite{Rubin1995} develops perturbation theory expressions for time
dependent dispersion coefficients in bimodal media.  \cite{Dagan2003}
and \cite{Fiori2003} study time-dependent apparent dispersion in a
similar bimodal setup as \cite{Eames1999} using a Lagrangian approach
combined with a self-consistent effective medium
approximation. \cite{Fiori2006}, \cite{Fiori2007} and
\cite{tyukhovactrw2016} study transport in composite media
characterized by Gaussian and non-Gaussian distributions of the
logarithm of hydraulic conductivity. \cite{Fiori2006} and
\cite{Fiori2007} derive semi-analytical expressions for particle
travel times in order to map the conductivity distribution on solute
breakthrough curves. \cite{tyukhovactrw2016} use a kinematic
relationship to relate the advection time over a single inclusion to
its conductivity as the basis for CTRW model to predict solute
breakthrough curves. While these approaches provide the methodology to
construct upscaled expressions for solute breakthrough curves, they do
not provide evolution equations for the average solute concentrations.

\cite{Silliman1987}, \cite{Murphy1997} and \cite{Levy2003Measurement} observed non-Fickian
behaviors for the breakthrough in tank experiment characterized by low
conductivity inclusions embedded in a sandy matrix. \cite{Berkowitz2000}
modeled the tailing behaviors observed in these experiments
 using a CTRW approach, whose parameters were estimated from the
 observed breakthrough curves. \cite{Ginn2001} use a
stochastic-convective streamtube approach to model aerobic
biodegradation in a column experiment with bimodal medium structure. 
\citet{Zinn2004} carried out experiments in tank experiments with
bimodal medium structure and derived an upscaled model to describe the observed breakthrough
curves. For the advectively dominated transport in background and
inclusions, the authors use a streamtube model, for
diffusion-dominated transport in the inclusion, a matrix diffusion
model. As shown in our paper, in the case of randomly distributed
inclusions, the streamtube model breaks down for large scale advective
transport because individual streamlines sample a random number of inclusions
that can be characterized by a Poisson distribution. 

In this paper we derive an upscaled model for advective transport in a
bimodal 2D medium with randomly placed circular inclusions.  The
methodology is based on a Lagrangian approach that allows to identify
and quantify the stochastic rules of advective particle motion in
disordered media. Similar approaches have been used in previous works
for the analysis and upscaling of pore-scale
transport~\citep{morales2017, Puyguiraud2019a} and for transport in
multi-Gaussian hydraulic conductivity fields~\citep{Hakoun2019} and
fractured media~\citep{Hyman2019}. Here, we use a Lagrangian approach
to gain understanding of the stochastic principles of transport in
random composite media through the analysis of advective trapping
events in low conductivity inclusions, and the distribution of flow
speeds sampled between them. This analysis facilitates the formulation
of upscaled transport as a multi-trapping model. This is considered a
first step towards a mobile-immobile mass transfer model of transport in
highly heterogeneous media that includes advection and diffusion in
the whole domain and that allows for parameter predictions based on a
given structure. In Section II we introduce the flow and transport
model used. In Section III we discuss the transport behavior of three
types of media: a single inclusion, a periodic packing of inclusions
and a random packing of inclusions. In Section IV we present the
upscaled model derived for the random packing and we give some
conclusions in Section V.
%
%
\section{Flow and transport model}\label{sec:flowtpt}
\begin{figure}
  \centering
  \includegraphics[width=.9\textwidth]{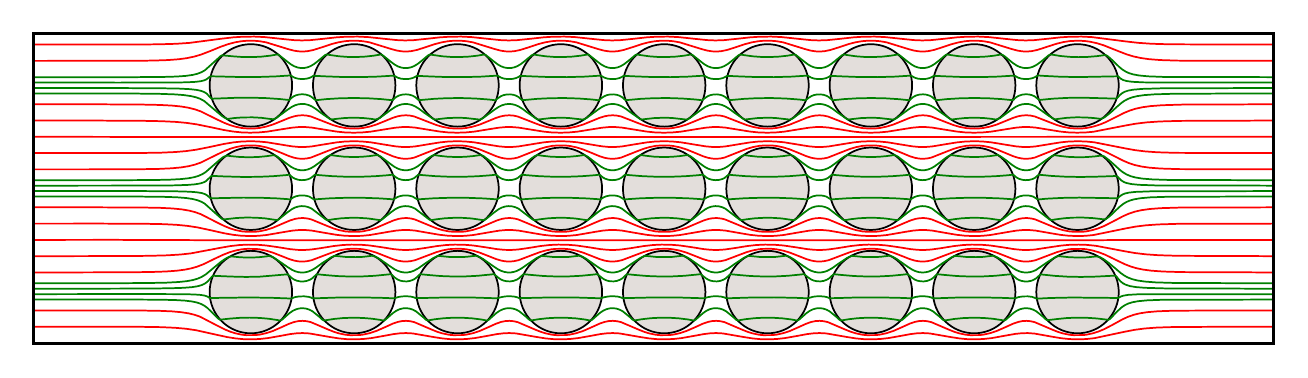}
  \includegraphics[width=.9\textwidth]{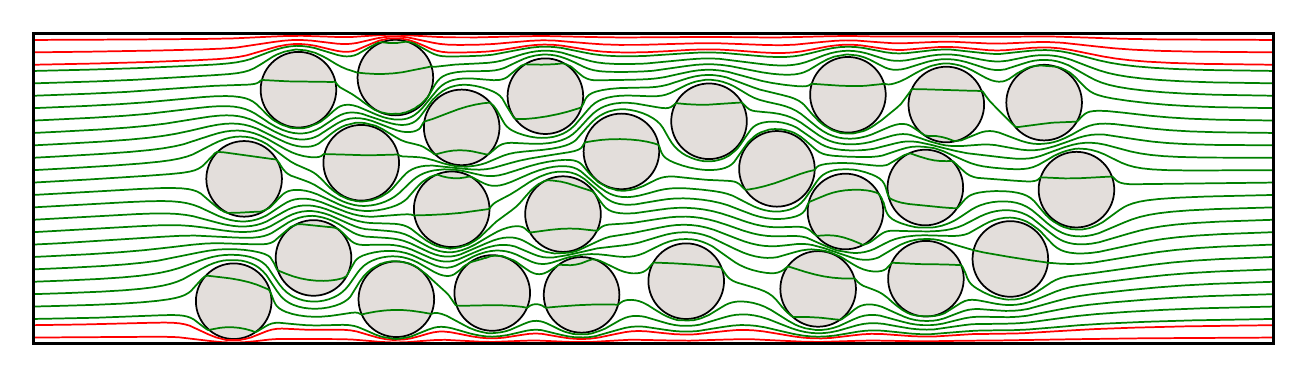}
  \caption{Flow and transport domain and streamlines of the Darcy flow
    $\mathbf q(\vx)$ for (top) regular and (bottom) random packing
    considering no flow boundary conditions at the top and bottom
    boundaries. Streamlines that cross at least one inclusion are
    green. Red streamlines do not go through any inclusion. In the
    regular media streamlines either cross all the inclusions in the
    horizontal or none of them. In the random media almost all
    streamlines cross at least on inclusion.  \label{fig:rnd}}
\end{figure}

We consider flow and transport in a 2D medium characterized by a
binary distribution of hydraulic permeabilities, where the material
with high permeability $K_m$ is connected (background material or
matrix), while the material with low permeability $K_i$ is
disconnected (inclusions). For simplicity we assume that the
inclusions have a circular shape of radius $r_{0}$ and can be
regularly or randomly arranged (see Figure~\ref{fig:rnd}).

Packings are characterized by the domain size $L_{x} \times L_{y}$,
inclusion radius, and covered volume fraction. In regular packings the
inclusions needed to cover the desired area are placed in a uniform
equispaced grid inside the domain (Figure~\ref{fig:rnd} top). Random
packings are generated by drawing the centers coordinates from an
uniform distribution and discarding inclusions that overlap previously
existing ones. The algorithm stops when the desired volume fraction is
covered. This method generates arrangements with an exponential
distribution of distances between inclusions (Figure~\ref{fig:rnd}
bottom).
%
%
\subsection{Flow}
We consider steady state flow through the medium described by the
Darcy equation
\begin{align}
  \label{eq:flow}
\mathbf q(\vx) = - K(\vx) \nabla h(\vx),
\end{align}
where $K(\vx)$ is the hydraulic conductivity, $h$ is the piezometric
head, and $\mathbf{q}$ is Darcy velocity. Both medium and fluid are
assumed to be incompressible, which implies that
$\nabla \cdot \mathbf q(\vx) = 0$. A constant flow rate $q_0$ is
imposed on the left domain boundary and constant head on the right one
so that the mean flow direction is along the $x$ axis.
%
%
\subsubsection{Flow distribution}
We discuss briefly here the flow distribution between the matrix and
the inclusions which will give us some information to analyze the
transport in the following sections.

For a single inclusion embedded in an infinite matrix, flow inside the
inclusion is uniform and aligned with the mean flow direction. The
ratio between undisturbed background flow velocity and the flow
velocity in the inclusion is given by~\citep{Wheatcraft1985}
\begin{align}
\label{eq:qin}
\frac{q_{in}}{q_0} = \frac{2 \kappa}{1+\kappa},
\end{align}
where $\kappa = K_i/K_m$ is the conductivity ratio.

For the composite media under consideration here, flow inside the
inclusions is in general not perfectly uniform and is not aligned with
the mean flow direction as shown in Figure~\ref{fig:rnd}. To estimate
the background flow velocity for small conductivity ratio under
consideration, flow through the inclusions may be disregarded compared
to the flow through the matrix. Thus, we can approximate the average
flow velocity in the background
\begin{align}\label{eq:vm-chi}
q_m = \frac{q_0}{1 - \chi}.
\end{align}
where $\chi$ denotes the volume fraction of the inclusions.
%
%
\subsection{Transport}
We consider purely advective transport, which is governed by the
following Liouville equation for the concentration $c(\vx,t)$
\begin{align}
  \label{eq:adv}
\frac{\partial c(\vx,t)}{\partial t} + \mathbf v(\vx) \cdot \nabla
  c(\vx,t) = 0,
\end{align}
where $\mathbf v(\vx) = \mathbf q(\vx) / \phi$. For simplicity,
porosity $\phi$ is assumed to be constant in this work. Note that this
is generally not true, particularly for geological media. However,
porosity in different materials varies typically between $0.05$ and
$0.4$, and this variation is much lower than that of hydraulic conductivity, which
may vary over several orders or magnitude between different
materials \citep{Bear1972}. Solute is initially uniformly distributed along a line
$c(\vx,t = 0) = c_0 \delta(x)$.

The transport problem is solved in a Lagrangian framework. The
equation of motion for the position $\vx(t;\mathbf a)$ of a fluid
particle is
\begin{align}
\frac{d \vx(t;\va)}{dt} = \mathbf v[\vx(t;\va)].
\end{align}
with $\vx(t=0;\va) = \va$. The distribution of initial positions is
$\rho(\va) = \delta(a_1)$. In the following transport will be
analyzed in terms of the arrival time distribution of particles at
increasing distances from the inlet.

For a medium with impermeable inclusions, macrotransport can be
described by the advection dispersion equation (ADE)
\begin{align} \label{eq:ADE}
\frac{\partial \overline c(x,t)}{\partial t} + v_a \frac{\partial
  \overline c(x,t)}{\partial x} - D_{a} \frac{\partial \overline
  c(x,t)}{\partial x^2} = 0,
\end{align}
where $v_a$ is the apparent velocity and the dispersion coefficient
$D_{a}$. For the condition of a low density of inclusions, i.e.
$\chi \ll 1$, \citet{Eames1999} report
\begin{align}\label{eq:EamesImpervious}
D_{a} = 0.74 \chi v_{0} r_{0}
\end{align}
for impermeable inclusions and
\begin{align}\label{eq:EamesLimitZero}
D_{a} = \frac{8}{3\upi\kappa} \chi v_{0} r_{0}
\end{align}
in the limit of $\kappa \to 0$.

The distribution of arrival times at a position $x_{c}$ for an
instantaneous injection into the flux at $x = 0$ is given
by~\cite{Ogata1961} and \cite{KreftZuber1978}
\begin{align}
\label{eq:f}
f(t,x_{c}) = \frac{x_{c} \exp\left[- (x_{c}-v_{a} t)^2/4 D_{a} t \right]}{\sqrt{4 D_{a} t^3}}.
\end{align}
For the complementary cumulative arrival time distribution, we obtain
accordingly
\begin{align}
\label{eq:ade_analytic}
  F(t,x_{c}) = \int\limits_t^\infty dt' f(t',x_{c}) = 1-\frac{1}{2} \left[\text{erfc}\left(\frac{x_{c} - v_{a} t}{\sqrt{4
  D_{a} t}}\right) + \exp\left(\frac{x_{c} v_{a}}{D_{a}}\right)\text{erfc}\left(\frac{x_{c}+v_{a}t}{\sqrt{4  D_{a} t}}\right)\right].
\end{align}
We use these solutions in the following as references for the observed
arrival time distributions. Furthermore, we estimate the apparent
velocity $v_{a}$ and apparent dispersion coefficient $D_{a}$ from the
mean $m_{b}$ and variance $\sigma_{b}^{2}$ of the
breakthrough time by using the Fickian relations
\begin{align}
\label{eq:vD}
v_{a} = \frac{x_{c}}{m_{b}}, && D_{a} = \frac{v_{a}^{3}\sigma_{b}^{2}}{2 x_{c}}.
\end{align}
%
%
%
\subsection{Numerical model}\label{sec:Numerics}
The rectangular domain of size $L_{x} \times L_{y}$ is discretized
using square cells of side $\Delta = r_{0}/30$. This discretization
ensures that the circular shape of the inclusions is well
reproduced. To avoid boundary effects the horizontal dimension is
extended a buffer length $\lceil 4r_{0} \rceil$ equally distributed
between the left and right boundaries.

Steady state flow \eqref{eq:flow} is solved using a two-point flux
finite volume scheme. Uniform velocity $v_{0}$ is prescribed on the
left boundary and head on the right boundary. The top and bottom
boundaries are periodic. Velocity is calculated on the cells sides.

The advection equation \eqref{eq:adv} is integrated using the
semi-analytical method of Pollock~\citep{Pollock1988}. At the beginning
of the simulation $N_{p} = 10^{6}$ particles are uniformly distributed
along the left boundary. The buffer between the boundary and the first
inclusions ensures that flow is uniform and the streamlines parallel
at the inlet. Therefore the uniform distribution of particles is
equivalent to a flux-weighted injection. The simulation runs until all
particles leave the domain. Streamlines and equivalently particle
trajectories are illustrated in Figure~\ref{fig:rnd}.

\begin{figure}
  \centering
   \includegraphics[width=.9\textwidth]{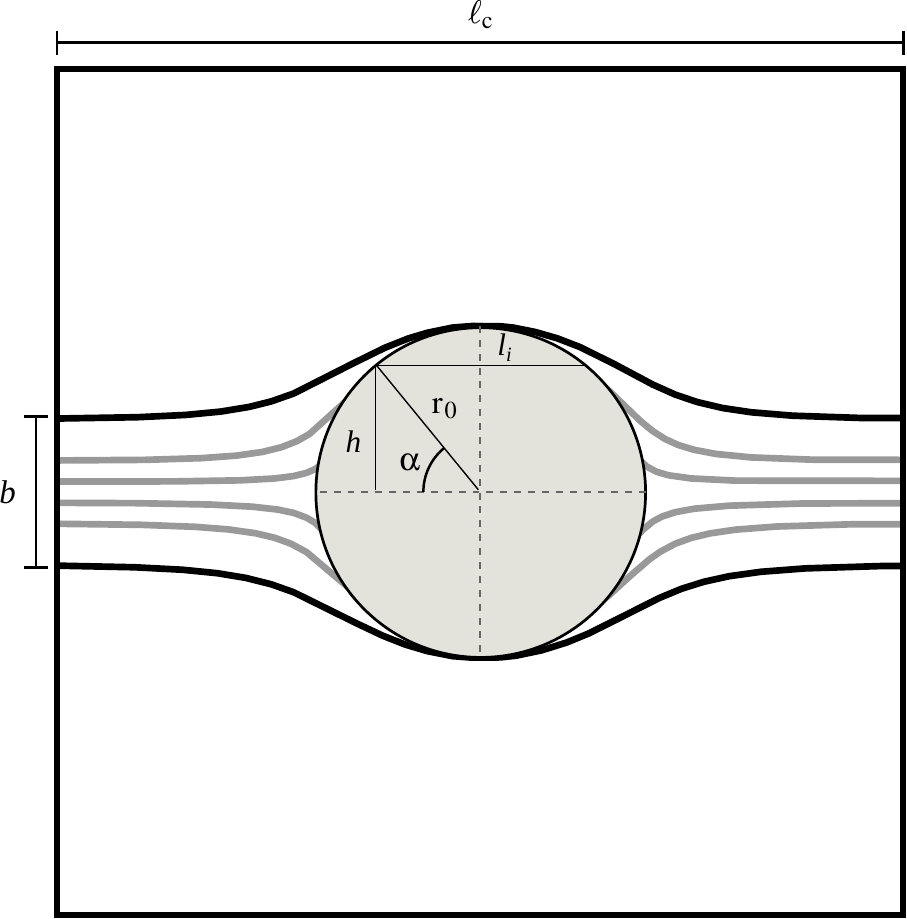}
   \caption{Sketch of the set up for a unit cell of size $\ell_{c}$
     single inclusion containing an inclusion of radius $r_{0}$. Only
     particles in the segment $b$ enter the inclusion. }
  \label{fig:inclusion}
\end{figure}

Results are reported in dimensionless units. We chose as
characteristic length the side of the unit cell $\ell_{c}$ of a
regular arrangement with inclusions of radius $r_{0}$ that covers a
volume fraction $\chi$ (Figure~\ref{fig:inclusion}). That is,
$\ell_{c}= r_{0}\sqrt{\upi/\chi}$. The characteristic time is
$\tau_{c} = \ell_{c}/v_{0}$ so that a dimensionless time of one is
required to traverse the unit cell at the prescribed velocity. The
time needed to travel through the buffer area is subtracted from the
results.
%
%
\section{Transport behavior}
We study the transport behavior in media with random arrangements of
inclusions. Transport is characterized by the travel time of the
particles in terms of the breakthrough curve or equivalently by the
complementary cumulative breakthrough curve at control planes. We will
also analyze the trapping events distribution (i. e., the number of
inclusions that a particle is transported through before arriving at
the control plane), and the velocity distribution inside the
inclusions. The velocity in the background material does not vary
much. However, the tortuosity of the flow paths leads to an enhanced
spreading of the particles as discussed for macrodispersion.

%
%
\subsection{Single inclusion}
We consider first the case of a medium in which there is only one
inclusion of low permeability (see Figure~\ref{fig:inclusion}). We
analyze the residence time of the particles within the inclusion and
the relation with the breakthrough curve.

\subsubsection{Residence times}
The residence time distribution in a single inclusion in an infinite domain is obtained by
purely geometrical considerations as follows. The flow field within an
isolated single inclusion is constant. Since the streamlines inside
the inclusion are parallel, the particles that go through it are
uniformly distributed over the vertical diameter. This means that the
vertical particle position is uniformly distributed in
$\left[-r_{0},r_{0}\right]$. The position $h$ of a particle on the
vertical diameter of the inclusion is $h = r_{0} \sin \alpha$. Therefore,
we obtain the angle $\alpha$ at which the particle entered the
inclusion as
\begin{align}
\alpha(h/r_{0}) = \arcsin{(h/r_{0})}.
\end{align}
From this we obtain the angular distribution that corresponds to the
uniform particle distribution as
\begin{align}
p_\alpha(\alpha) = \cos{\alpha}.
\end{align}
The length of the circle segment traversed by the particle $l_i$ is given by
$s(\alpha) = 2 r_{0} \cos{\alpha}$, whose distribution $p_{l_{i}}(l_{i})$
is obtained from $p_\alpha(\alpha)$ as
\begin{align}
  p_{l_{i}}(l_{i}) = p_\alpha[\arccos{(l_{i}/2r_{0})}]
  \left.\frac{1}{dl_{i}(\alpha)/d\alpha}\right|_{\alpha = \arccos{(l_{i}/2r_{0})}}
\end{align}
Thus,
\begin{align}
\label{eq:p_li}
p_{l_{i}}(l_{i}) = \frac{l_{i}}{r_{0}} \frac{1}{\sqrt{1 - (l_{i}/2r_{0})^2}}
\end{align}
and the distribution of transition times $t = l_{i}/v_{in}$ is given by
\begin{align}
  \label{eq:1incl-res-times}
  \psi(t) = \frac{t}{\tau_{in}^2} \frac{1}{\sqrt{1 - (t / \tau_{in})^2}},
\end{align}
where we defined $\tau_{in} = 2r_{0}/v_{in}$ is the maximum advection
time across the inclusion. The comparison
between~\eqref{eq:1incl-res-times} and residence times obtained
numerically is shown in Figure~{\ref{fig:1incl-res-times}.
\begin{figure}
  \centering
  \includegraphics[width=.9\textwidth]{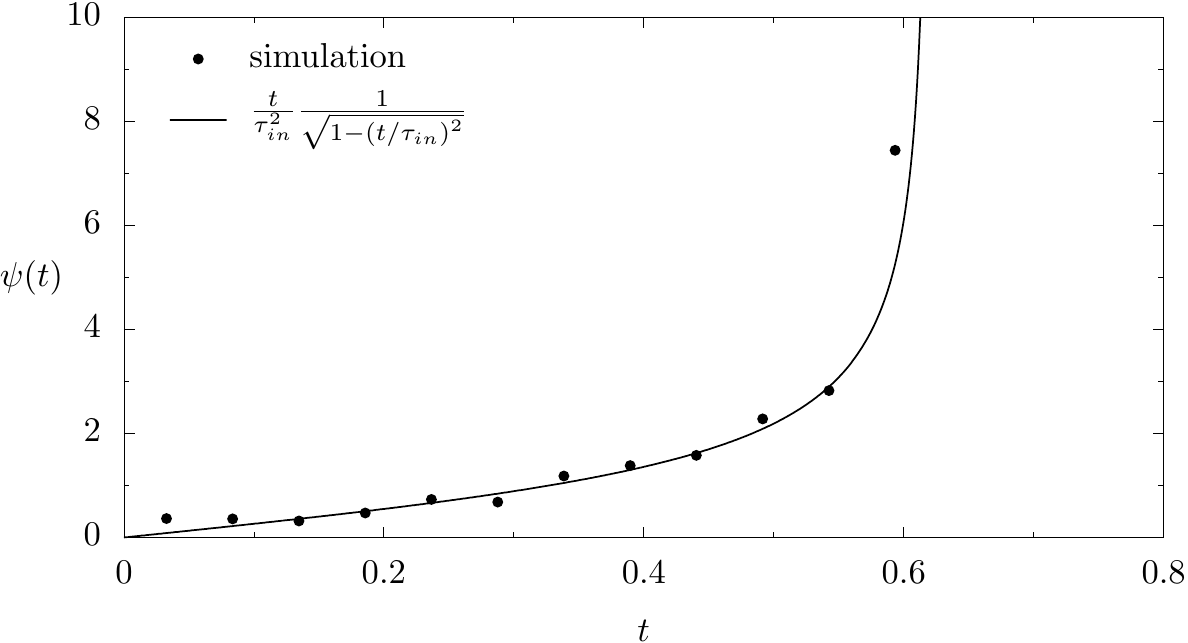}
  \caption{Comparison between the simulated (dots) and analytical
    (solid line) residence time distribution of particles traveling
    through a single inclusion $\chi =0.01$, and $\kappa = 0.1$.
    \eqref{eq:1incl-res-times} }\label{fig:1incl-res-times}
\end{figure}
%
%
\subsubsection{Fraction of particles traversing the inclusion}
The fraction of particles entering over a length $\ell_c$ that traverses through the inclusion is obtained
from flux conservation. The size $b$ of the streamtube in the matrix
passing through the inclusion is obtained from
\begin{align}
2 r_{0} v_{in} = b v_m
\end{align}
Thus, the flux proportion that goes through the inclusion can be
written as
\begin{align}
\label{eq:a0}
a_0 = \frac{b}{\ell_{c}} = \frac{2 r_0 v_{in}}{\ell_{c}v_m} = \frac{4 r_0}{\ell_c} \frac{\kappa}{1 + \kappa},
\end{align}
where we used expression~\eqref{eq:qin}.
\subsubsection{Breakthrough curves}
The breakthrough and the complementary cumulative breakthrough curves
for the single inclusion are shown in Figure~\ref{fig:1incl-cbtc}. The
first arrival occurs at $t=1$, which
correspond to the time needed to go through the unit cell. Part of the
streamlines are bent by the presence of the low permeability inclusion
causing the peak to widen. The rest of the curve reflects the effect
of the low permeability inclusion with a breakthrough curve
(Figure~\ref{fig:1incl-cbtc}~right) that follows the above calculated
residence time distribution.
\begin{figure}
  \centering
  \includegraphics[width=.45\textwidth]{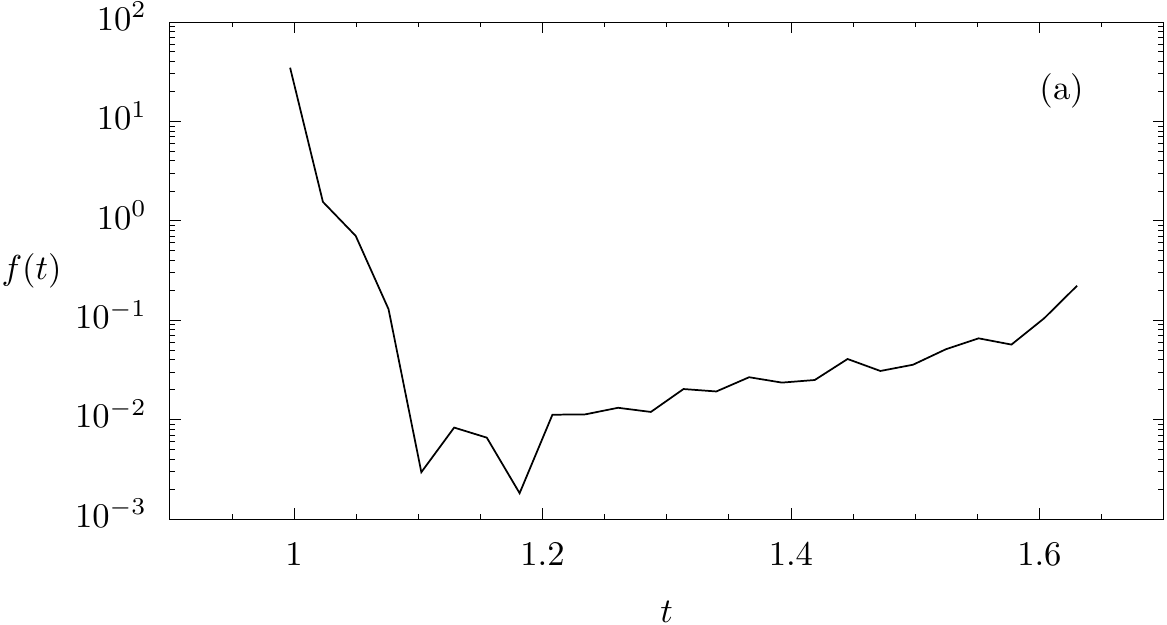}
  \includegraphics[width=.45\textwidth]{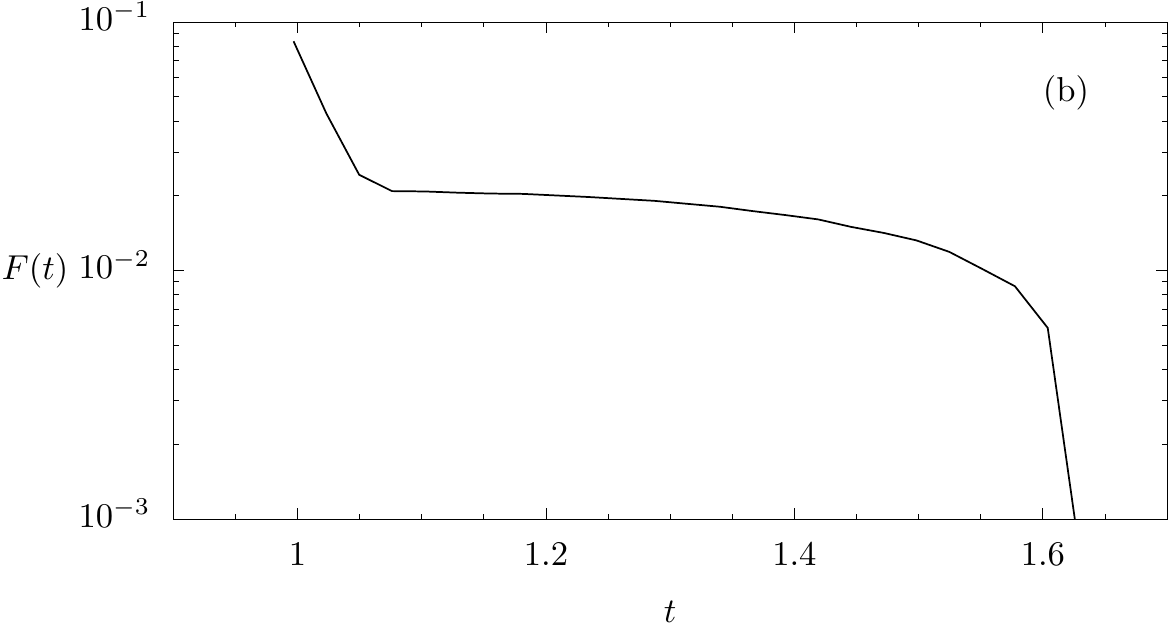}
  \caption{Breakthrough curve (a) and complementary cumulative
    breakthrough curve (b) for a system containing one inclusion
    ($\chi =0.01$, $\kappa = 0.1$) measured at a distance $\ell_{c}$
    from the inlet. The curves show a early arrival of particles that
    only travel through the matrix and a long tail formed by the
    particles traversing the inclusion at different
    heights.}\label{fig:1incl-cbtc}
\end{figure}

Transport through a single inclusion can be conceptualized as a
streamtube model with two types streamtubes. In one of them, a
percentage of particles $a_{0}$~\eqref{eq:a0} is transported
through the inclusion, while in the other one particles are
transported only through the matrix. This conceptual model can be
extended to regular packings whose unit cell contains only one
inclusion. In this case particles will either travel through all the
inclusions in the streamtube or none of them (see
Figure~~\ref{fig:rnd}). The travel times inside of each streamtube are
distributed due to the tortuosity of the streamlines. For regular
packings the streamlines differ from the single inclusion case because
of the finite size of the unit cell, which enforces a straight
streamline at its boundary.

Based on the conceptual model of two streamtubes and considering that
the inclusions are much less permeable than the background, the
breakthrough curve is characterized by two distinct pulses caused by
transport in the streamtubes without and with inclusions
(Figure~\ref{fig:1incl-cbtc}). Note that the transition is continuous,
as the outermost streamlines of the two streamtubes coincide .

For short media, the periodic medium could be a useful approximation
to predict breakthrough curves. \citet{Zinn2004} carried out
experiments of solute transport in two-dimensional glass bead packs,
where circular inclusions were randomly placed into a less permeable
background. They used the streamtube approach to predict breakthrough
curves for the advective dominated case. As the approximate solution
is based on one single inclusion, periodicity is inherently
assumed. In the measured breakthrough curves, the double breakthrough
behaviour is very clear and it could be demonstrated that their
streamtube approach worked well to reproduce the breakthrough curves
(see also next subsection).
%
%
\subsection{Random packings}\label{sec:rp}
We consider now random packings of inclusions generated as explained
in Section~\ref{sec:flowtpt}. First we consider media of different
sizes ($3 \le L_{x}/\ell_{c} \le 500$;
$1 \le L_{y}/\ell_{c} \leq 105$) and covered volume fraction
($0.1 \le \chi \leq 0.55$) in which we study the velocity distribution
in the matrix and inclusions, the trapping events experienced by
particles and the trapping time distribution.

\begin{figure}
  \centering
  \includegraphics[width=.9\textwidth]{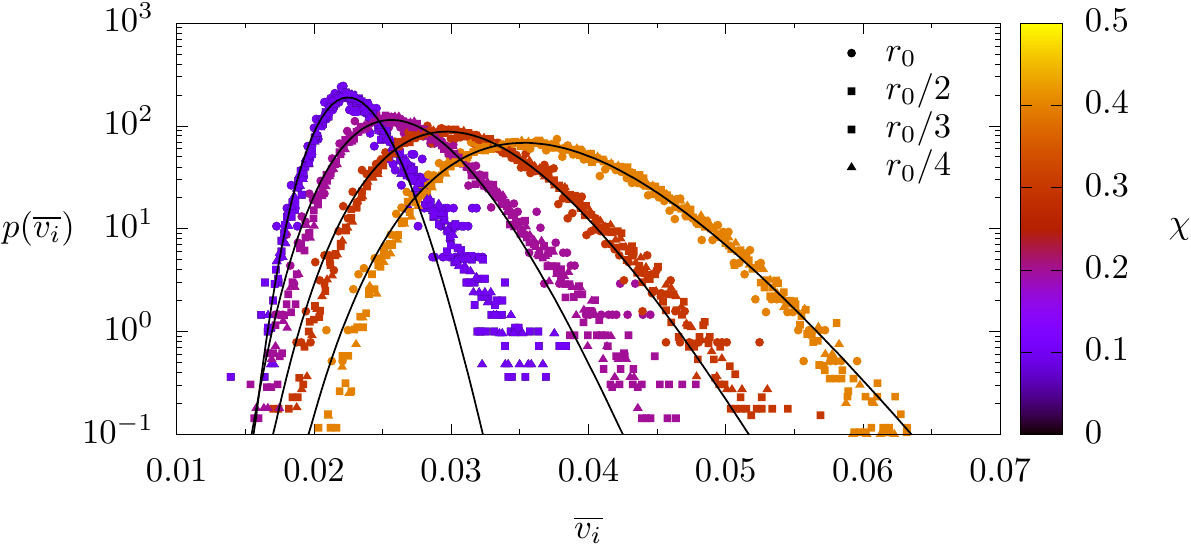}
  \caption{Mean velocity distribution inside the inclusions (symbols)
    for a media with varying volume fraction and inclusions' size with
    constant $\kappa = 0.01$. The solid lines show the fit to a
    log-normal distribution to the all the data with same $\chi$. The
    base case geometry is $L_{x}=49.7\ell_{c}$, $L_{y}=2.5\ell_{c}$,
    and $\chi=0.1$. The rest of the cases keep the same domain
    proportions. }\label{fig:vel-dist}
\end{figure}

Then we study the behavior of breakthrough curves. First we explore
further the streamtube model using the geometry
of~\citet{Zinn2004}. Next we consider two scenarios, a long medium
($387\ell_{c} \times 3.8\ell_{c}$) in which transport is analyzed as
the distance from the inlet increases, and a wide medium
($84\ell_{c} \times 28\ell_{c}$) in which the effect of the length of
the line along which solute is injected is studied.
%
%
\subsubsection{Velocity distribution\label{sec:vpdf}}
The velocity inside isolated regularly arranged inclusions is
approximately constant under low density of inclusions conditions, this means for $\chi
\ll 1$. For increasing $\chi$ in random packing this is in general not
the case and flow velocities vary inside the inclusions and between
inclusions.
We characterize the inclusions by their mean velocities, and study
their distributions $p_v(v)$ as a function of volume fraction
$\chi$. Figure~\ref{fig:vel-dist} shows the distributions of inclusion
velocities for different volume fractions and inclusion sizes.  We
observe that the distribution of mean velocities can be well
approximated by a log-normal distribution. Consistent with
\eqref{eq:qin} and \eqref{eq:vm-chi}, the mean velocity of the
distribution is independent of the inclusion size and depends only on
the volume fraction $\chi$ for constant $\kappa$. The distribution
becomes narrower with decreasing $\chi$. In fact, in the limit of low
density of inclusions, $p_v(v)$ should converge to the Delta
distribution $p_v(v) = \delta(v-v_{in})$, where $v_{in}$ is the
constant velocity in a single isolated inclusion.

The velocity in the matrix (Figure~\ref{fig:vel-mat}) is inversely
proportional to the covered area ratio $\chi$ and follows the relation
\eqref{eq:vm-chi} until a high volume fraction is covered, reaching
the percolation threshold, and the hypothesis that flow through the
inclusions is negligible compared to the flow through the matrix is no
longer valid.
\begin{figure}
  \centering
  \includegraphics[width=.9\textwidth]{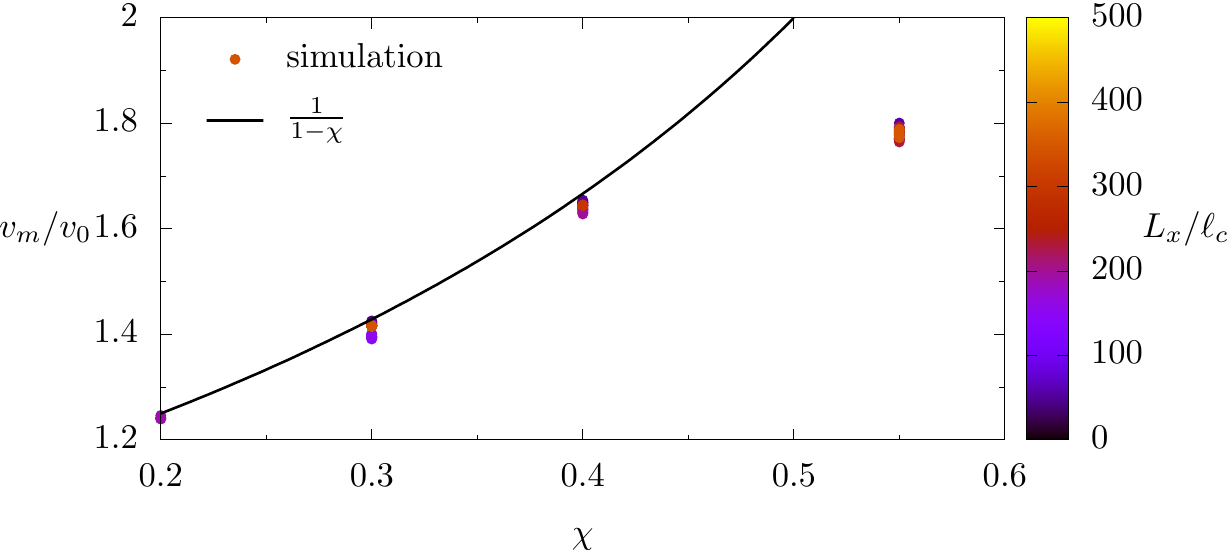}
  \caption{Mean velocity in the matrix versus volume fraction occupied
    by the inclusions. The solid line correspond to the
    solution~\eqref{eq:vm-chi} for isolated inclusions. Dots colors
    correspond to the medium length.}\label{fig:vel-mat}
\end{figure}
%
%
\subsubsection{Trapping events}
The number of trapping events experienced by a particle in random
media is not binary distributed as in the regular ones. As the
inclusions are randomly distributed in space, the distance between them
is approximately an exponential distribution, or in other words, the
number of inclusions that may be encountered within a given distance
follows a Poisson process~\citep{Feller1968}. In fact, we find that
the statistics of the number $n_{tr}$ of trapping events within a
travel distance $\ell$ can be described by the Poisson distribution
\begin{align}\label{eq:trap-poisson}
  p(n_{tr},\ell) = \frac{e^{-k\ell}\left(k \ell\right)^{n_{tr}}}{n_{tr}!}.
\end{align}

An example of the trapping events distributions is shown in
Figures~\ref{fig:trap-events-Lx}. It can be seen that the distribution
of trapping events evolves as the particles sample the medium. At
short distance from the inlet the distribution is narrow suggesting
that for a small medium the streamtube approximation could be
sufficient to explain transport. As the distance from the inlet
increases, the distribution widens and the probability of not being
trapped decreases. At a sufficient travel distance all particles
experience at least one trapping event and the distribution converges
to a Poisson distribution.
\begin{figure}
  \centering
  \includegraphics[width=.9\textwidth]{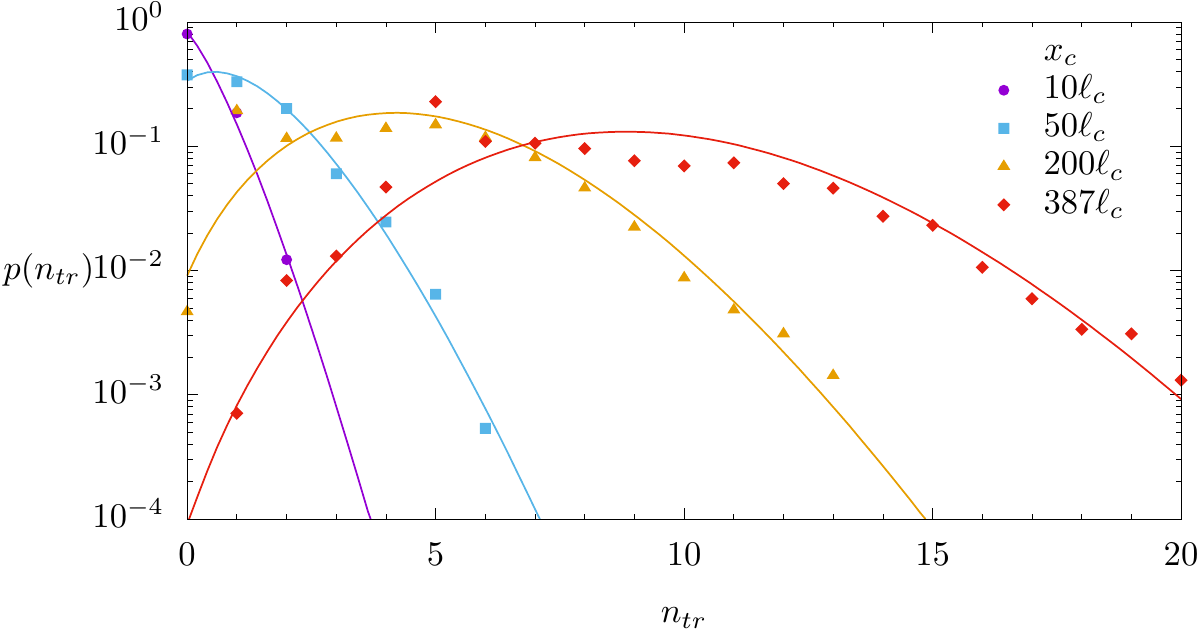}
  \caption{Distribution of number of trapping events (symbols) at
    different distances $x_{c}$ from the inlet for an arrangement of
    inclusions ($L_{x}=387\ell_{c}$, $L_{y}=3.87\ell_{c}$,
    $\kappa = 0.01$, $\chi=0.3$; same as in Figure
    \ref{fig:rndLx}). The solid lines are the fit to a Poisson
    distribution.}\label{fig:trap-events-Lx}
\end{figure}

The trapping rate $k$, that is the number of trapping events per
traveled distance, that characterizes the Poisson distribution depends
on the geometry of the arrangement. To assess this dependence we
performed a series of simulations varying the medium geometry (radius,
length, width, and area covered by the inclusions). The average
distance between the inclusions $d$ was computed with the following
algorithm. First, we take the lines between all pairs of inclusions'
centers that do not intersect another inclusion. Then, for every pair
of lines that intersect, the shortest one is kept. Finally, the
average length of the remaining lines is calculated. As shown in
Figure~\ref{fig:trap-freq} the trapping rate is inversely proportional
to the average distance between the inclusion $d$, expressed in terms
of the unit cell size $\ell_{c}$.
\begin{figure}
  \centering
  \includegraphics[width=.9\textwidth]{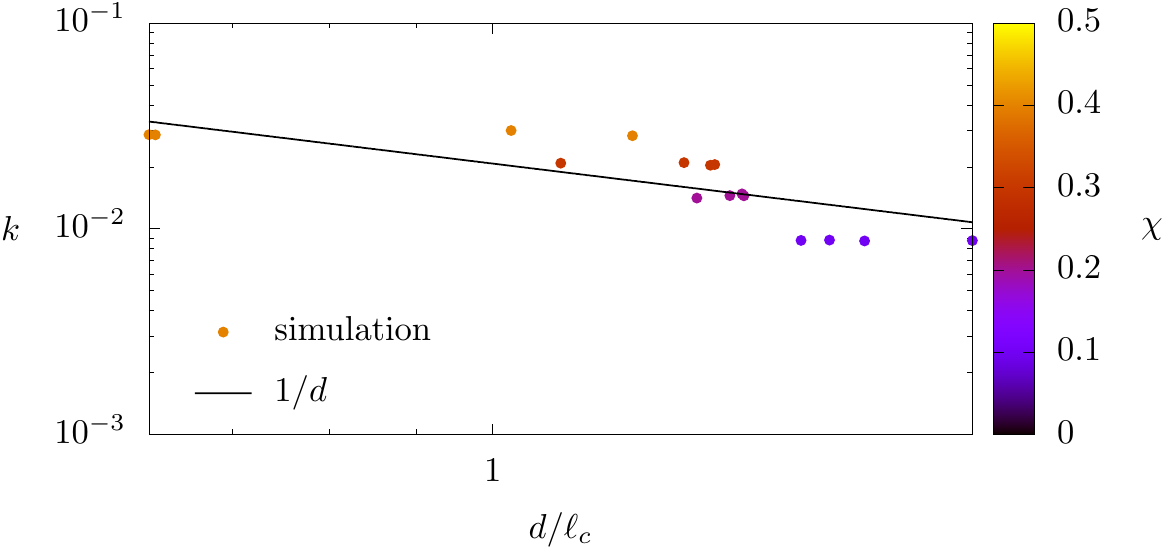}
  \caption{Trapping frequency (parameter in Poisson distribution)
    versus mean distance between inclusions. Point color is the
    covered volume fraction $\chi$.}\label{fig:trap-freq}
\end{figure}
%
%
\subsubsection{Distribution of trapping times}
The trapping time distribution is obtained numerically from the residence time
distribution in a single inclusion \eqref{eq:1incl-res-times}. The distribution of
trapping times in the following is denoted by $\psi_{f}(t)$. It can
be constructed from the distribution $\psi(t|v)$ of trapping times for
a given inclusion velocity, and the distribution $p_v(v)$ of
velocities as
\begin{align}
\label{eq:traptimesdist}
\psi_f(t) = \int\limits_0^\infty d v p_v(v) \psi(t|v).
\end{align}
Figure~\ref{fig:trap-times} compares the distribution of trapping
times obtained from the direct numerical simulations and the
model~\eqref{eq:traptimesdist}.
\begin{figure}
  \centering
  \includegraphics[width=.9\textwidth]{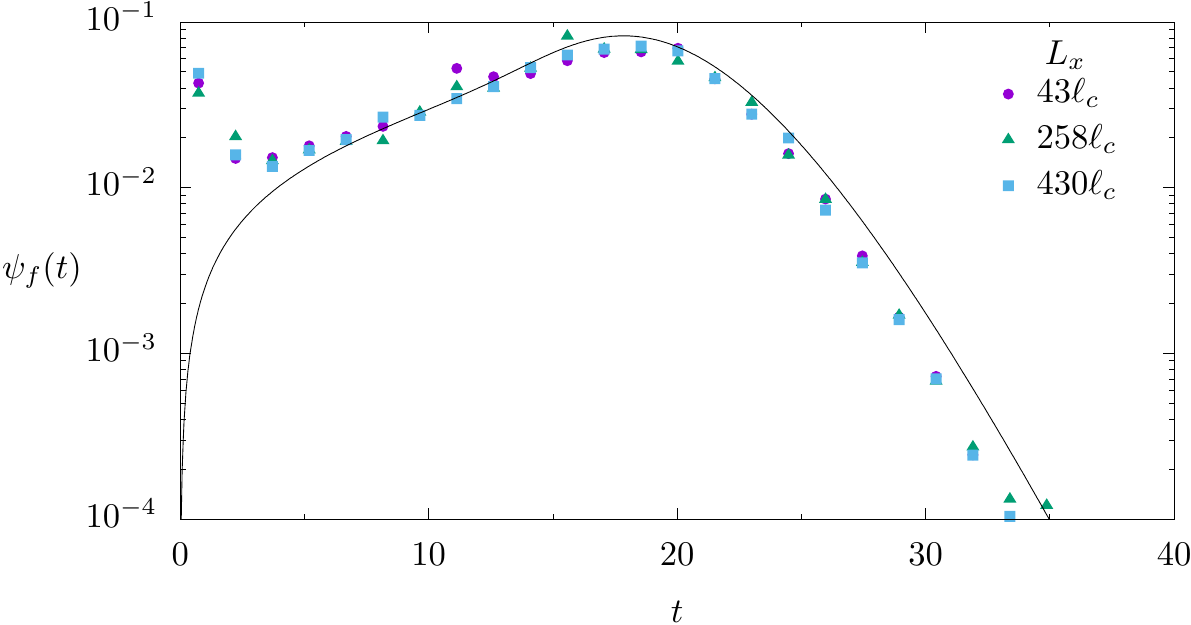}
  \caption{Comparison between the theoretical trapping times
    distribution given by~\eqref{eq:traptimesdist} (black line) and
    the numerical results for media with $L_{y}=3.87\ell_{c}$,
    $\kappa = 0.01$, $\chi=0.3$, and different length
    (symbols).}\label{fig:trap-times}
\end{figure}
%
%
\subsubsection{Breakthrough curves}
We consider first a medium based on the inclusions geometry of
\citet{Zinn2004}. This medium has $\chi = 0.37$,
$L_{x}= 10.8\ell_{c}$, and $L_{y} = 5.4\ell_{c}$. We consider an
intermediate permeability ratio scenario with $\kappa= 0.01$. The
breakthrough curve (Figure~\ref{fig:zinn-btc}) is affected by the
random arrangement of inclusions. However, we can distinguish the
contribution of particles that experience different numbers of trapping
events. Given the small size of the domain and the low number of
inclusions, particles experience only a few trapping events and most
of them travel through the domain without entering any
inclusion. Based on this phenomenology, \citet{Zinn2004} used a
streamtube approach in order to model the breakthrough curves observed
in their experiment. Their approach identified a streamtube passing
only through the matrix and a second streamtube that passes through a
constant number of inclusion. This approach is not valid in a large
medium characterized by a random arrangement of inclusions because
streamlines  may pass through random numbers of inclusions, as
discussed below.
\begin{figure}
  \centering
  \includegraphics[width=.49\textwidth]{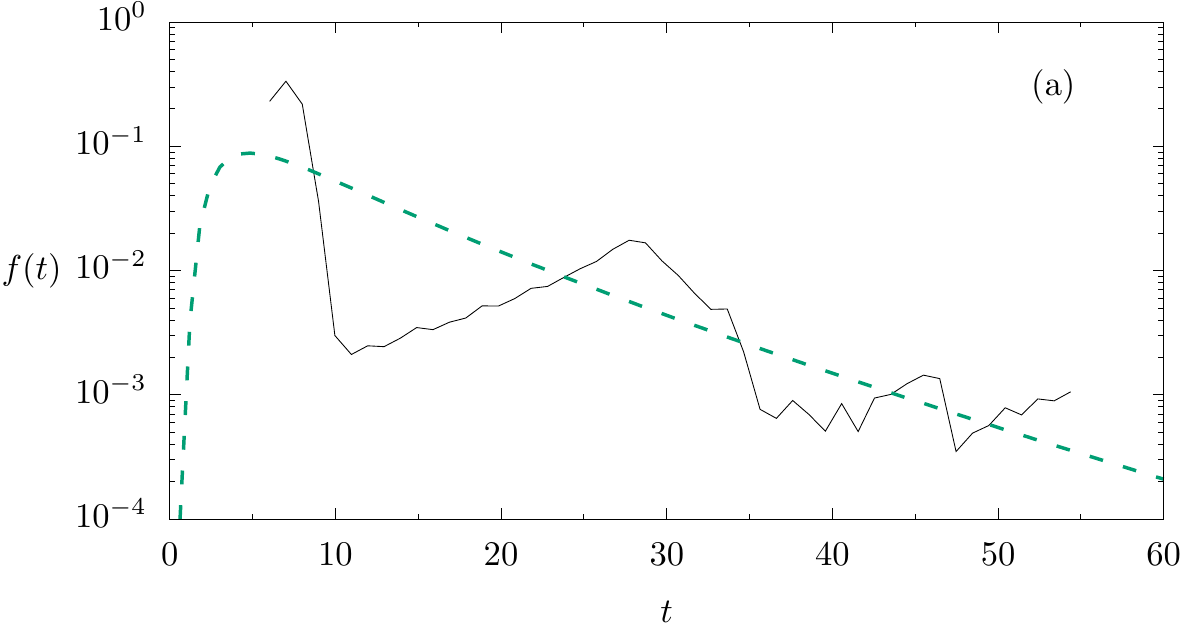}
  \includegraphics[width=.49\textwidth]{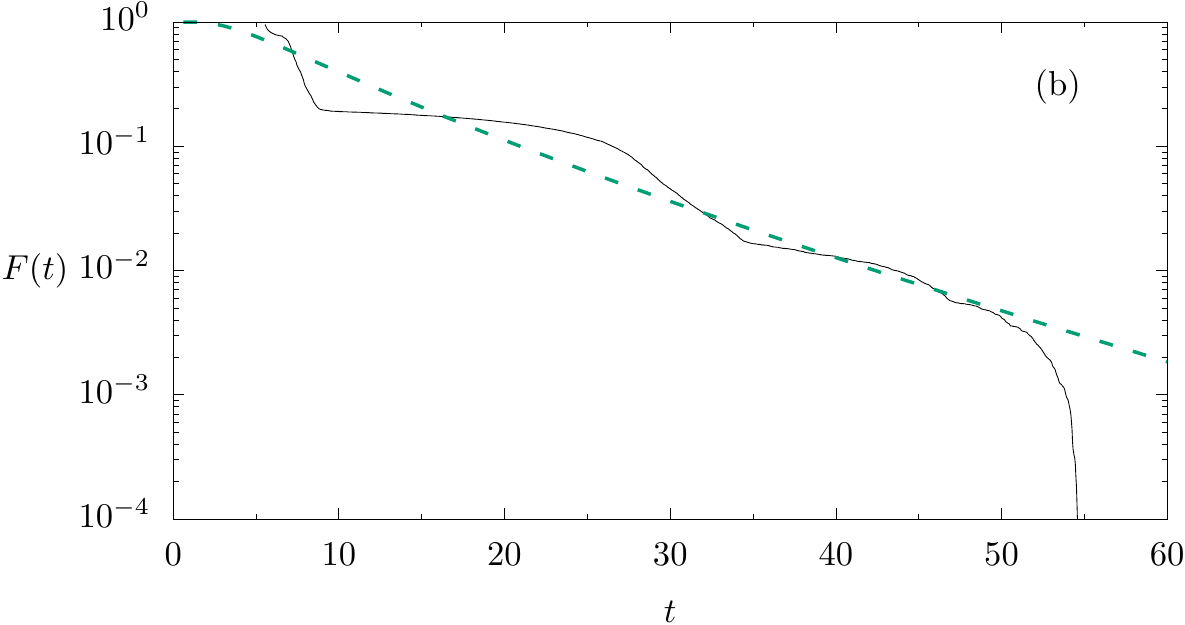}
  \caption{Breakthrough curve (a) and complementary cumulative
    breakthrough curve for a random medium with a geometry as in
    \citet{Zinn2004} ($L_{x}= 10.8\ell_{c}$, $L_{y} = 5.4\ell_{c}$,
    $\kappa= 0.01$, and $\chi = 0.37$). The dashed lines correspond to
    the analytical solution~\eqref{eq:ade_analytic} where
    $D_{a}$ and $v_{a}$ are obtained from the mean and variance
    of the breakthrough time.}\label{fig:zinn-btc}
\end{figure}

Next we consider a long and narrow medium
($387\ell_{c} \times 3.8\ell_{c}$), where particles can travel through
a larger number of inclusions. As shown in Figure~\ref{fig:rndLx} the
shape of the breakthrough curves depends on the traveled distance,
that is, the amount of medium heterogeneity sampled. The curves become
smoother as the distance from the inlet increases. For short distances
(Figure~\ref{fig:rndLx}~a and d) the first part of the curve is dominated by
the dispersion caused between the streamlines along the fast paths and
the tail of the curve by the streamlines going through the inclusions
as in the case of the geometry of \citet{Zinn2004}. For a sufficiently
long distance from the inlet (Figure~\ref{fig:rndLx}~b,~c, e, f), the
shapes of the breakthrough curves suggest that
the peak and tail behavior can be modeled by an effective hydrodynamic
dispersion coefficient. The parameters of the apparent center of mass
velocity and dispersion coefficients are obtained from the
breakthrough data according to~\eqref{eq:vD}. Their values are given
in Table~\ref{tab:velDisp}. 

The average velocity fluctuates little,
and is close or equal to the velocity set by the flow boundary condition. The
dispersion coefficient is variable and evolves with
distance from the inlet plane. The corresponding Fickian
solutions~\eqref{eq:f} and ~\eqref{eq:ade_analytic} provide good
descriptions of the breakthrough curves at large distances
($x_{c} > 300 \ell_{c}$, Fig.~\ref{fig:rndLx}c and f) from the inlet plane. However, the dispersion
coefficients differ from the ones obtained by~\citet{Eames1999} for
impermeable inclusions \eqref{eq:EamesImpervious}, $D_{a}= 0.069$, and
in the limit $\kappa \to 0$ \eqref{eq:EamesLimitZero}, $D_{a} =
7.87$. As pointed out by~\citet{Eames1999}, their expressions are
valid in the low density of inclusions limit of $\chi \ll 1$, which is
not the case for the volume fractions under consideration here. The
fact that the inclusion velocities are distributed, as discussed in
Section~\ref{sec:vpdf}, is a manifestation of the interaction between
inclusions, this means, they cannot be considered isolated.

In summary, while the Fickian solution fits the peaks and part of the tails at
intermediate and large distances from the inlet plane ($x_c > 50 \ell_c$), it fails to
reproduce the sharp cut-offs at early and late times, and completely
fails to reproduce the breakthrough curves at short distances ($x_c <
10 \ell_c$). Furthermore, the apparent dispersion coefficients fitted to the data
evolve with distance from the inlet plane, which cannot be accommodated by a standard
Fickian model based on constant transport parameters.

%
\begin{figure}
  \centering
  \includegraphics[width=.45\textwidth]{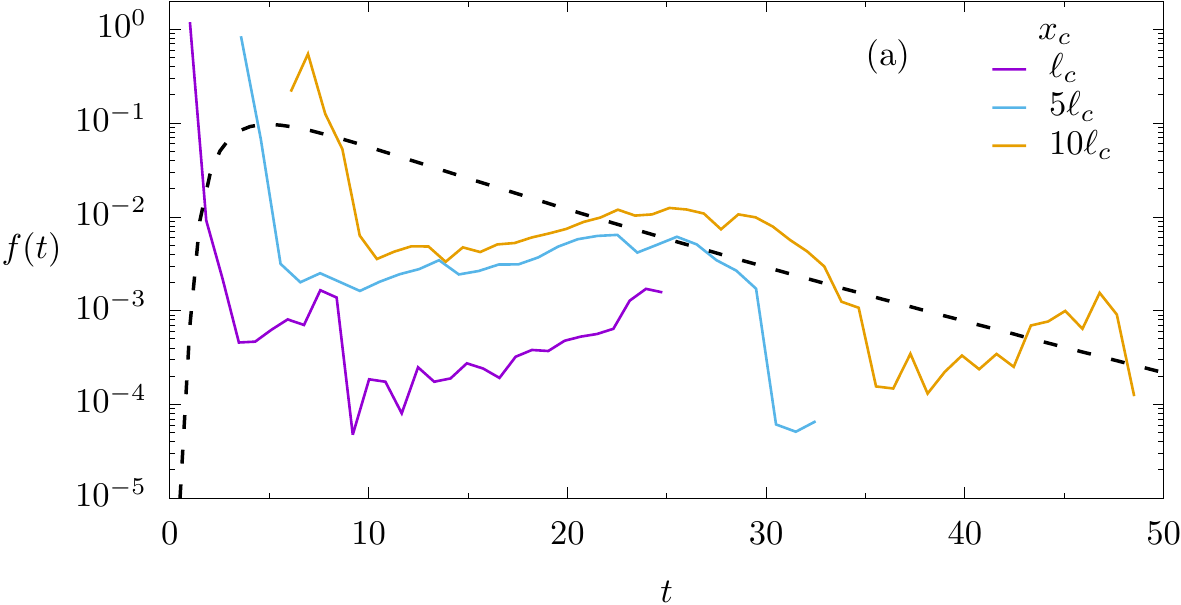}
  \includegraphics[width=.45\textwidth]{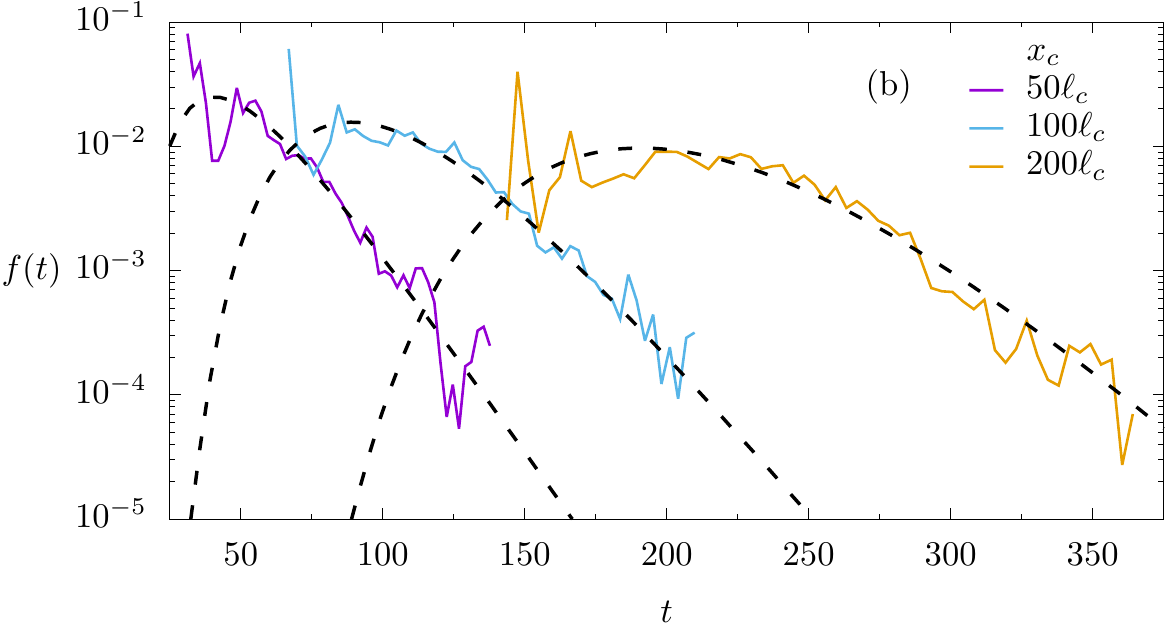}
  \includegraphics[width=.45\textwidth]{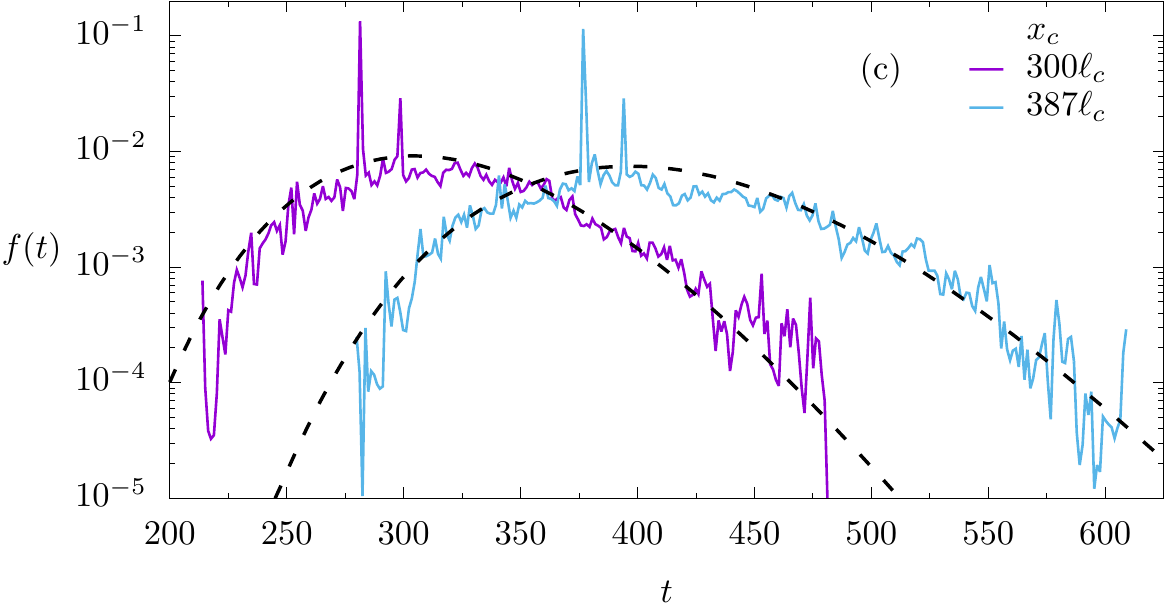}
  \includegraphics[width=.45\textwidth]{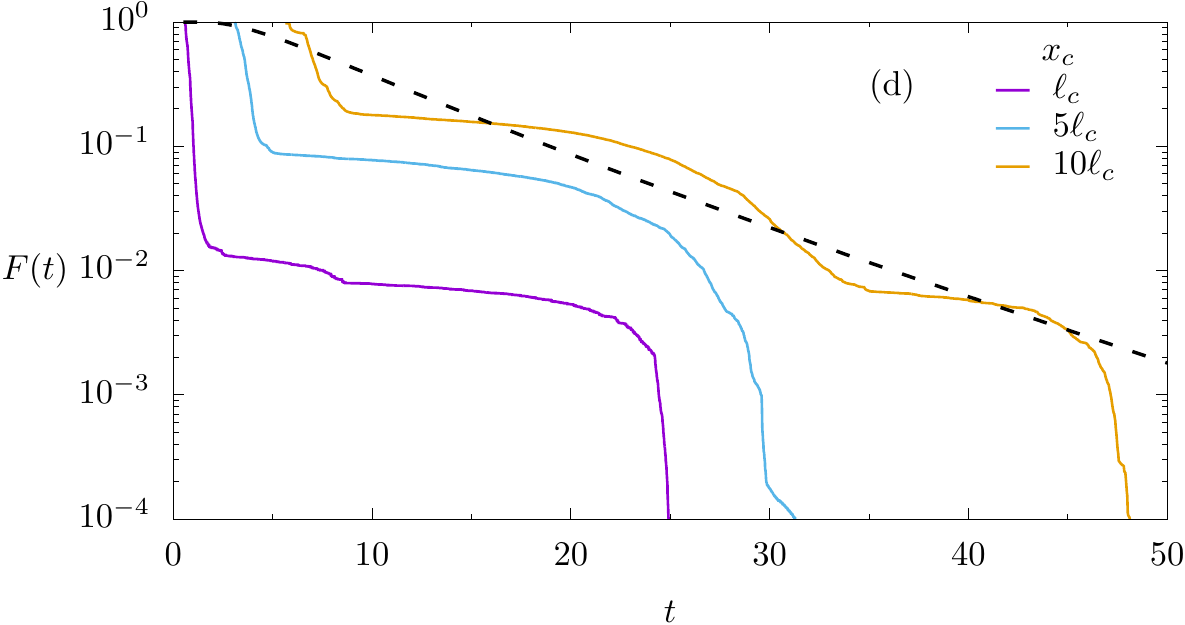}
  \includegraphics[width=.45\textwidth]{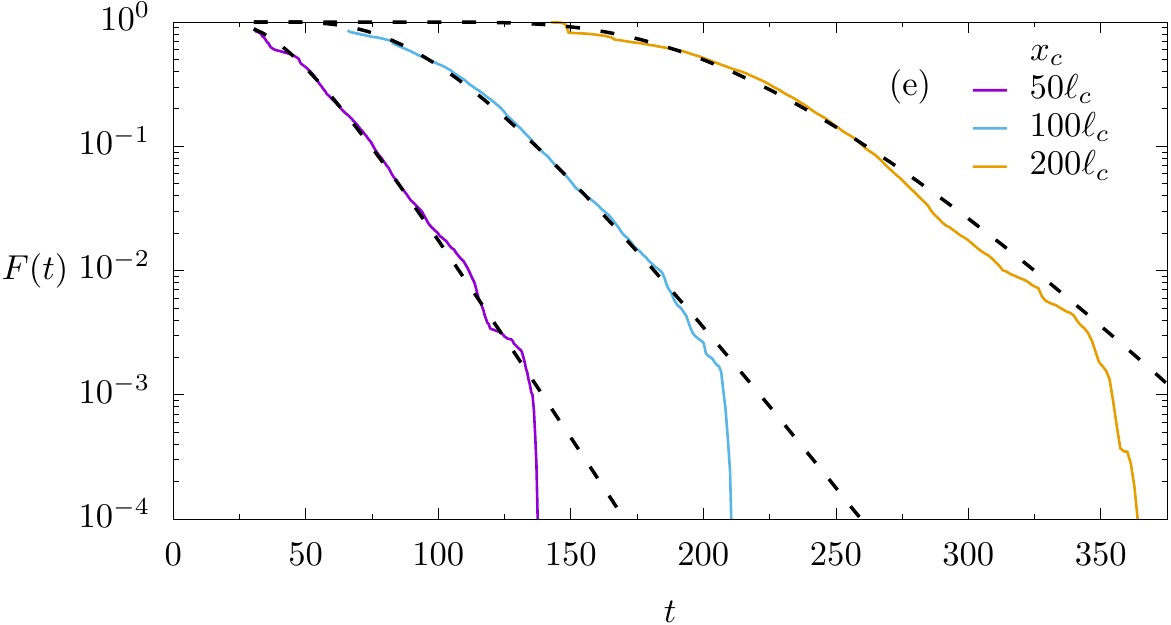}
  \includegraphics[width=.45\textwidth]{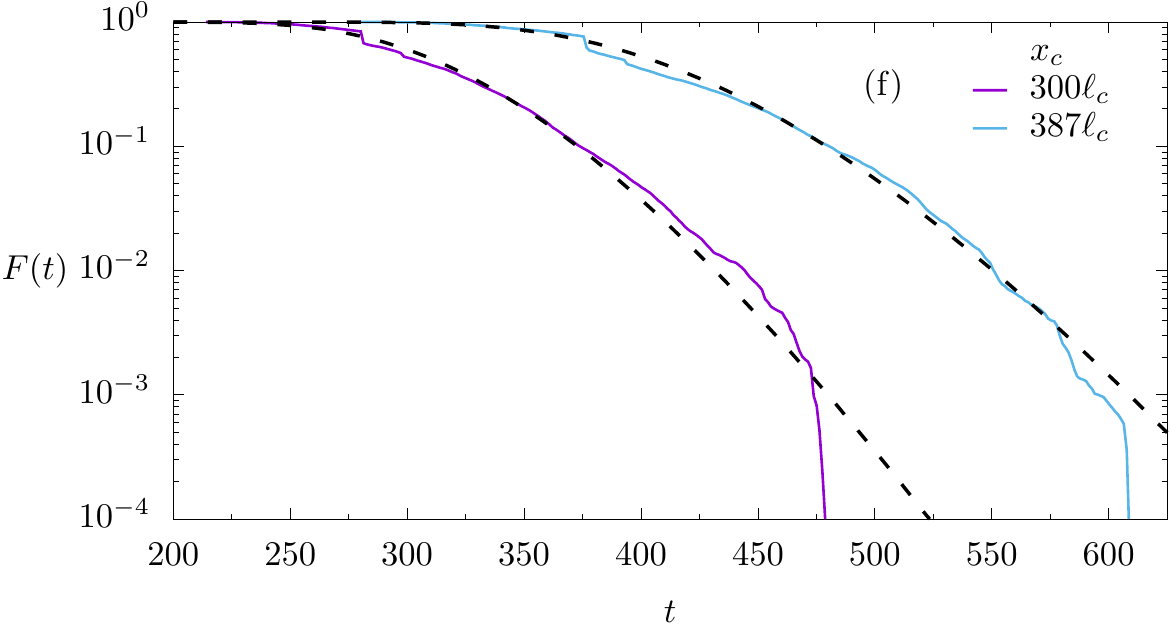}
  \caption{Breakthrough curves (a--c) and complementary cumulative
    breakthrough curves (d--f) at different distances from the inlet
    for a random arrangement of inclusions ($L_{x}=387\ell_{c}$,
    $L_{y}=3.87\ell_{c}$, $\kappa = 0.01$, $\chi=0.3$). The dashed
    lines correspond to the analytical
    solution~\eqref{eq:ade_analytic} where $D_{a}$ and $v_{a}$ are
    given in Table~\ref{tab:velDisp}. Note that the cases
    $ x_{c} = \ell_{c}, 5\ell_{c}$ are not
    modelled. }\label{fig:rndLx}
\end{figure}
\begin{table}
  \begin{center}
    \def~{\hphantom{0}}
    \begin{tabular}{llllllll}
      $x_{c}$ & 10 & 50 & 100& 200 & 300 & 387 \\
      $v_{a}$ &~1  &~1 &~~1 &~~0.98 &~~0.96 &~~0.96\\
      $D_{a}$ &~2.59 &~3.48 &~~3.87&~~4.42&~~2.86 &~~3.28\\

    \end{tabular}
  \end{center}
  \caption{Values of velocity $v_{a}$ and dispersion coefficient
    $D_{a}$ are determined from the mean and variance of the
    corresponding breakthrough times. For comparison the values
    predicted by \citet{Eames1999} are $D_{a}= 0.0686$ for impermeable
    inclusions and $D_{a} = 7.87$ for $\kappa \ll 1$.} \label{tab:velDisp}
\end{table}

In order to study the impact of the width of the initial particle
distribution on heterogeneity sampling we performed another series of
simulations in a wide medium ($84\ell_{c} \times 28\ell_{c}$) in which
we considered injection lines of increasing length (from
$0.28\ell_{c}$ to the whole width of the medium; centred at $L_{y}/2$)
and computed the breakthrough curves at different distances from the
inlet. The breakthrough curves (Figure~\ref{fig:InjSize}) show that
injection lines of small length (Figure~\ref{fig:InjSize}a,b with
injection lines of length $0.28\ell_{c}$ and $1\ell_{c}$ respectively)
do not sample enough of the medium variability even at the maximum 
travelled distance simulated. The curves have distinct peaks/bumps,
whose number increases with the distance as the number of trapping
events experienced by the particles grows. This behavior is similar to
the streamtube behavior observed for the long medium
(Figure~\ref{fig:rndLx}) and in the geometry of \citet{Zinn2004}
(Figure~\ref{fig:zinn-btc}). For injection lines of length above
$5\ell_{C}$ (Figure~\ref{fig:InjSize}c,d), the medium properties are
better sampled and the shape of the curves are more similar between
injections. The shape and number of peaks/bumps is less dependent on
the travelled distance and only the tail of the curve changes. Note
that the medium is not long enough to observe the asymptotic behavior
of Figure~\ref{fig:rndLx} d.
\begin{figure}
  \centering
  \includegraphics[width=.45\textwidth]{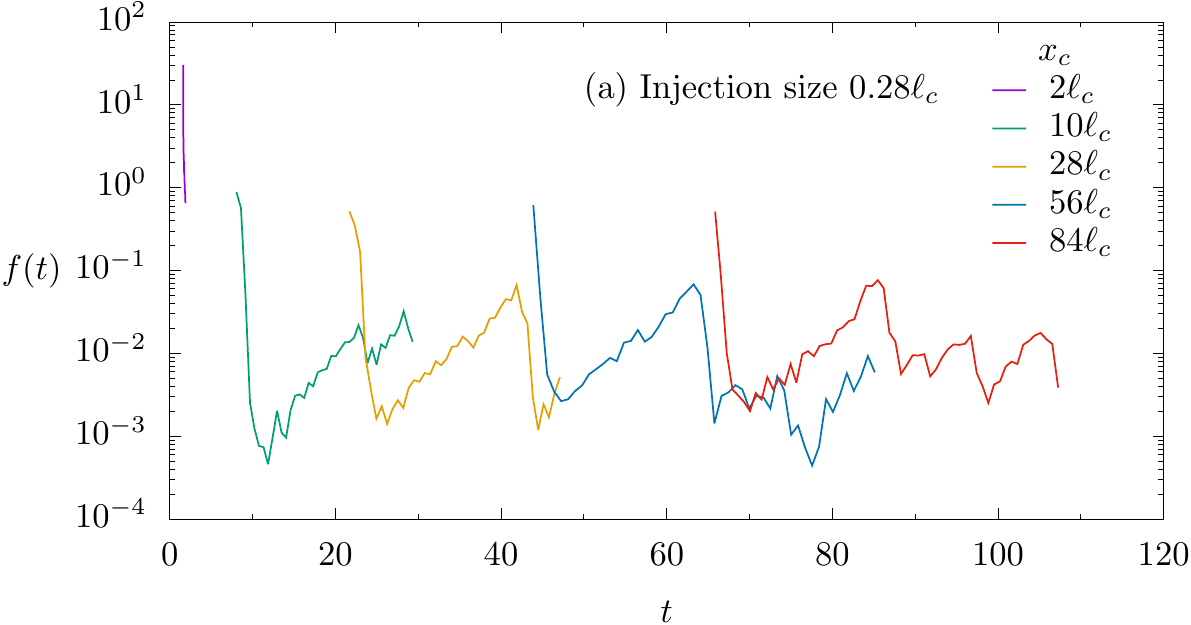}
  \includegraphics[width=.45\textwidth]{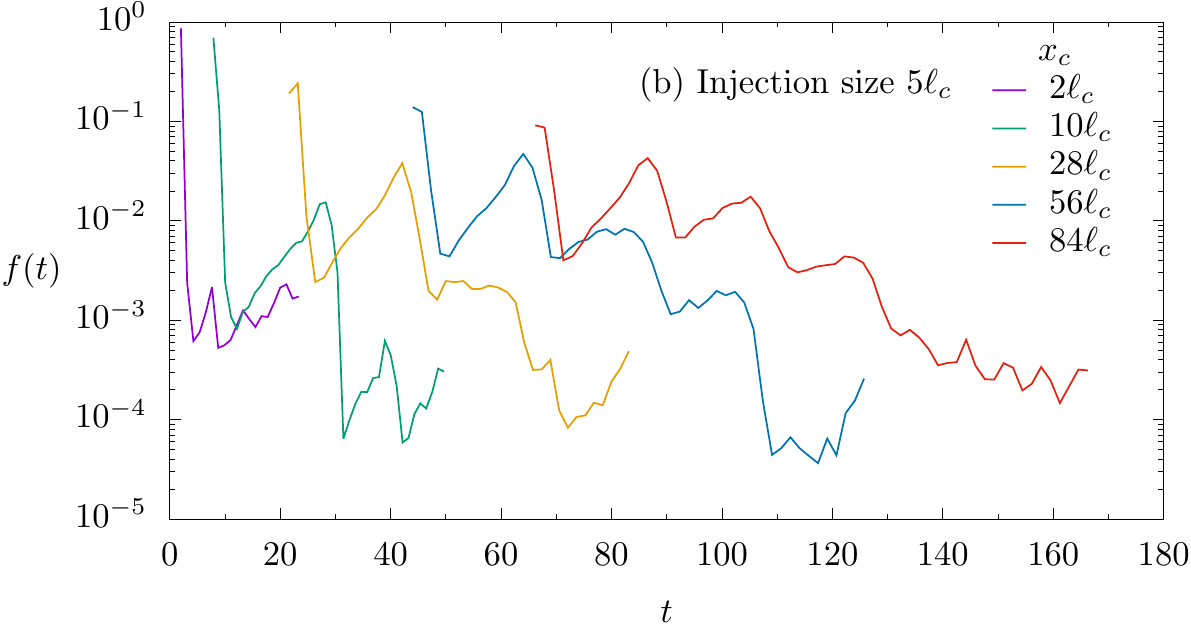}
  \includegraphics[width=.45\textwidth]{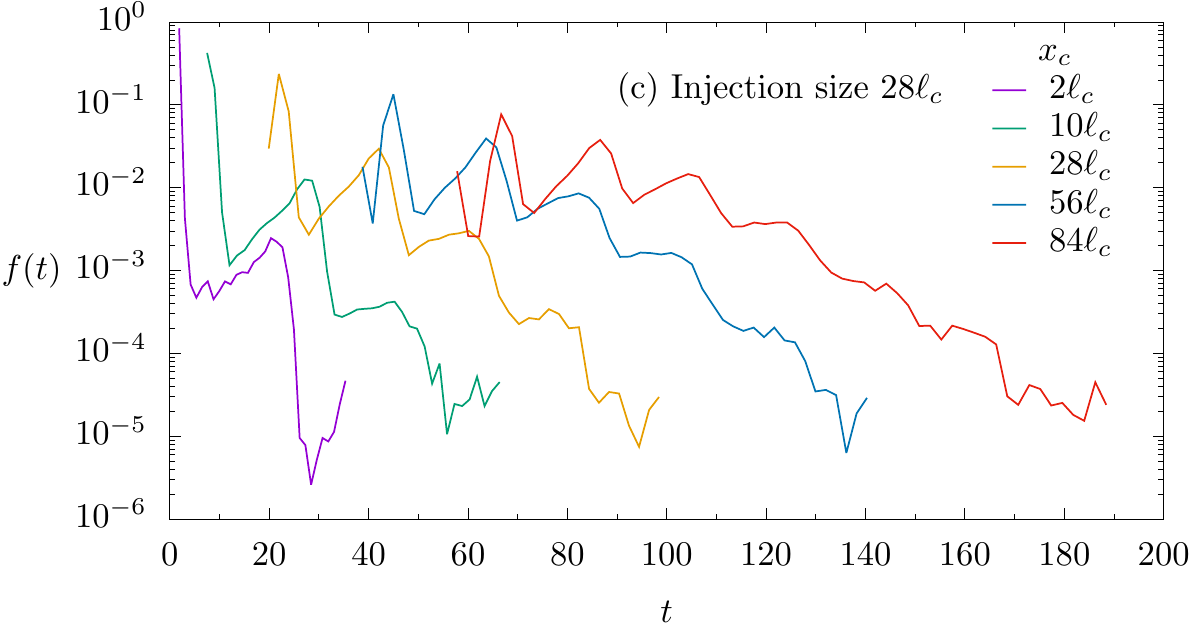}
  \includegraphics[width=.45\textwidth]{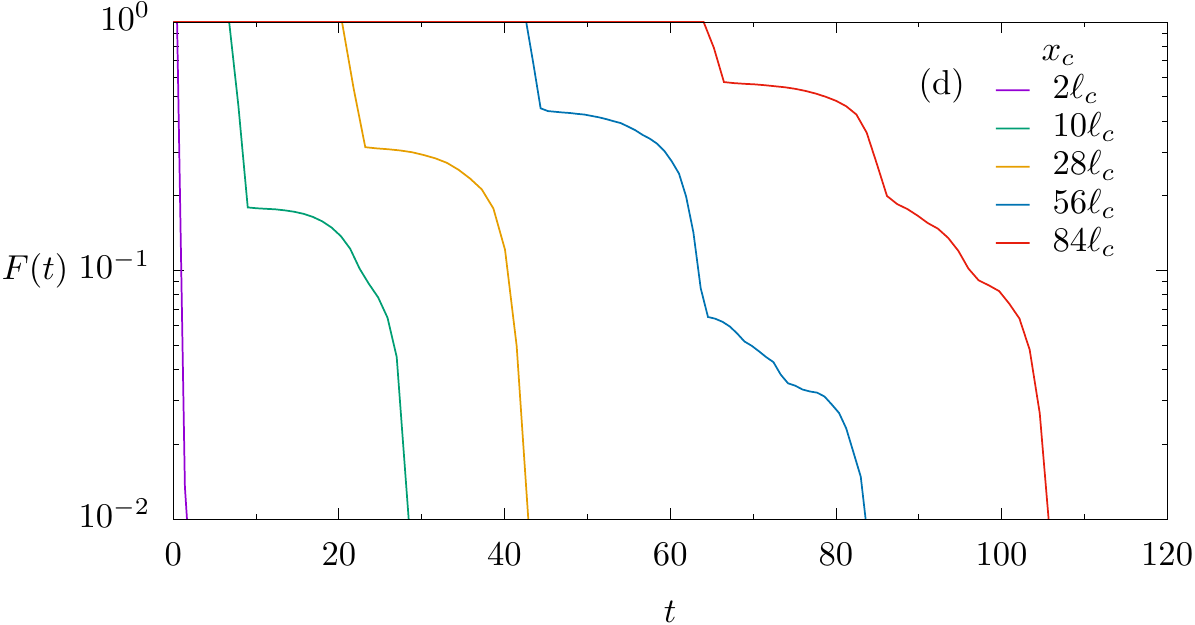}
  \includegraphics[width=.45\textwidth]{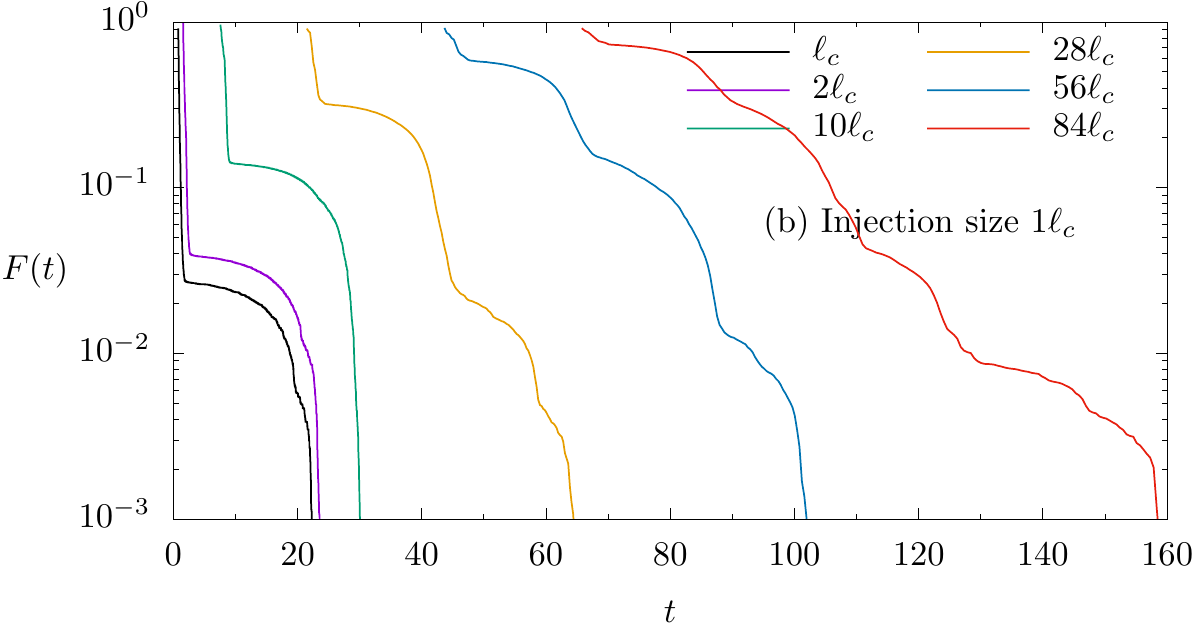}
  \includegraphics[width=.45\textwidth]{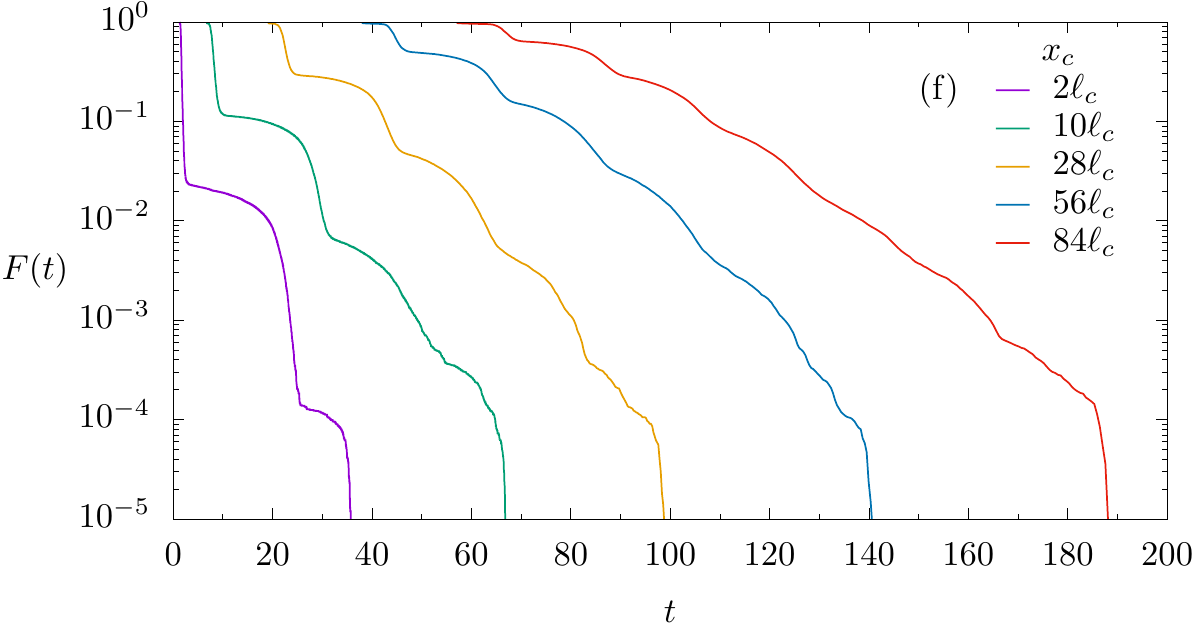}
  \caption{Breakthrough curves at different distances from the inlet
    for injection lines of increasing length (a, d $0.28\ell_{c}$, b, e
     $5\ell_{c}$, and c, d $28\ell_{c}$) for a medium with
    $L_{x} = 84\ell_{c}$, $L_{y} = 28\ell_{c}$, $\kappa = 0.01$, and
    $\chi=0.2$).}\label{fig:InjSize}
\end{figure}
%
%
\section{Upscaled transport model}
We derive an upscaled model for transport through random
packings. Unlike regular packings, in which streamtubes traverse
either the same number of inclusions or none of them, in random
packings streamtubes can sample a random number of inclusions. This
means, one cannot distinguish only two kinds of streamtubes, but one
has a set of streamtubes, each of which is characterized by different
random numbers of inclusions. The analysis of Section~\ref{sec:rp} has shown
that the number of trapping events, this means, the number of
inclusions a particle crosses along a trajectory, can be represented
by the Poisson distribution~\eqref{eq:trap-poisson} characterized by the
trapping rate $k$.

Based on these observations, we can now quantify the upscaled particle
motion using a continuous time random walk (CTRW)
framework~\citep{Berkowitz2006, noetinger2016}. To this end, we
consider advective-dispersive particle transitions in the mobile
matrix
\begin{subequations}
\label{eq:ctrw}
\begin{align}
\label{ctrw}
dx(s) = v_{m} ds +\sqrt{2D_m ds} \xi(s),
\end{align}
where $s$ denotes the mobile time spend outside the inclusions,
$v_{m}$ is the mean velocity in the matrix, $D_{m}$ is the
longitudinal dispersion coefficient, $\xi(s)$ is a Gaussian white
noise characterized by zero mean and unit variance. In the model,
$v_{m}$ can be estimated from the covered area $\chi$
(Figure~\ref{fig:vel-mat}) and $D_{m}$ is taken equal to the dispersion
coefficient for impermeable inclusions
\eqref{eq:EamesImpervious}. During the mobile time $s$ particles
encounter $n_{s}$ inclusions, where $n_{s}$ is distributed according
to~\eqref{eq:trap-poisson}. The clock time $t(s)$ after the mobile
time $s$ has passed is given by
\begin{align}
\label{eq:time}
t(s) = s + \sum\limits_{i = 1}^{n_{s}} \tau_{i}
\end{align}
where $n_{s}$ is Poisson distributed with mean value
$\langle n_{s} \rangle = k v_{m} s$. The trapping times $\tau_{i}$ are
defined by (see Appendix~\ref{app:upscaling})
\begin{align}
\tau_{i} =  \frac{\ell_{i}}{v_{i}} - \frac{\ell_{i}}{v_{m}},
\end{align}
where the distance $\ell_{i}$ is the secant of the circular inclusion
at the height where the particle enters the inclusion
(Figure~\ref{fig:inclusion}). It is distributed according
to~\eqref{eq:p_li}. The velocity $v_{i}$ in the inclusion is assumed
to be constant and lognormally distributed (see Section~\ref{sec:vpdf}
and Figure~\ref{fig:vel-dist}). The trapping time denotes the time a
particle spends in the inclusion minus the time it would take to
traverse the inclusion with the mean velocity $v_{m}$. Thus it
quantifies the net impact of the inclusion. The
medium is considered ergodic if each particle samples the same
distribution $\psi_{f}(t)$ of trapping times as it moves through the
medium. This property is clearly not fulfilled for a periodic medium
and depends on the medium and injection line length in random media.
According to the above, the clock time $t(s)$ is a compound Poisson
process~\citep{Feller1968,Margolin:et:al:2003,BensonMeer2009,Comolli2016}. Thus, its distribution $\psi(t)$ can be written in
Laplace space as (see also Appendix~\ref{app:upscaling})
\begin{align}
\label{eq:psi}
\psi^{\ast}(\lambda|s) = \exp\left(- \lambda s  - k s v_m \left[1 -
  \psi^{\ast}_{f}(\lambda) \right] \right),
\end{align}
\end{subequations}
where Laplace transformed quantities are marked by an asterisk, and
$\lambda$ denotes the Laplace
variable. Equations~\eqref{ctrw}--\eqref{eq:psi} constitute and
upscaled CTRW model combined with a multi-trapping approach. In the
following, we discuss the equivalent formulation in terms of a
time non-local partial-differential equation that describes advective
mobile-immobile mass transfer.  

In Appendix~\ref{app:upscaling}, we
derive for the mobile, this means non-trapped, solute concentration
$c_m(x,t)$ the governing equation
\begin{subequations}
\label{eq:mrmt-all}
\begin{align}
\label{eq:mrmt}
\frac{\partial c_{m}(x,t)}{\partial t} + \frac{\partial}{\partial t}
  \int\limits_{0}^{t} dt' \varphi(t - t') \gamma  c_{m}(x,t') + v_{m} \frac{\partial c_{m}(x,t)}{\partial x} - D_{m} \frac{\partial^{2} c_{m}(x,t)}{\partial x^{2}} = 0,
\end{align}
where the trapping rate is given by $\gamma = k v_m$. The
memory function $\varphi(t)$ is given explicitly in terms of the
advective trapping time distribution $\psi_{f}(t)$ as
\begin{align}
\label{eq:phi}
\varphi(t) = \int\limits_{t}^{\infty} dt' \psi_{f}(t').
\end{align}
The trapping time distribution $\psi_{f}(t)$, defined in Eq.~\eqref{eq:traptimesdist}, is determined by
the inclusion size and flow velocities within the inclusions. This
means, it is fully quantified in terms of the microscopic advective
trapping mechanisms. The memory function~\eqref{eq:phi} denotes the probability that the
trapping time is larger than the time $t$. Thus, we can define the
immobile concentration $c_{im}(x,t)$ as
\begin{align}
\label{eq:cimm}
c_{im}(x,t) = \int\limits_{0}^{t} dt' \varphi(t - t')  \gamma c_m(x,t'),
\end{align}
\end{subequations}
This equation reads as follows. The immobile concentration is
equal to the probability that a particle gets trapped in the immobile
region at any time $t' < t$ times the probability that the trapping
time is smaller than $t - t'$. Note that in the special case of a
single advection time scale $\tau_a$, this means for $\psi_f(t) =
\delta(t - \tau_a)$, the memory function~\eqref{eq:phi} reduces to a
step function as considered in~\cite{Ginn2017}. 

The upscaled model defined by~\eqref{eq:mrmt}--\eqref{eq:cimm} is
equal in form to memory function formulations of (multirate)
mobile-immobile mass transfer~\citep{Haggerty1995, Carrera1998,
  Schumeretal2003, Dentz2003, Ginn2017}, and defines the immobile
concentration in terms of the memory function $\varphi(t)$ as
expressed by~\eqref{eq:cimm}. The memory function here quantifies
advective mass transfer between high and low conductivity regions, and
is fully defined in terms of the advective trapping mechanisms,
inclusion size and velocity distribution. The
formulation~\eqref{eq:mrmt}--\eqref{eq:cimm} of the upscaled model in
terms of the non-local partial differential equation can be considered
as an advective mobile-immobile mass transfer model.
\begin{figure}
  \centering
  \includegraphics[width=.45\textwidth]{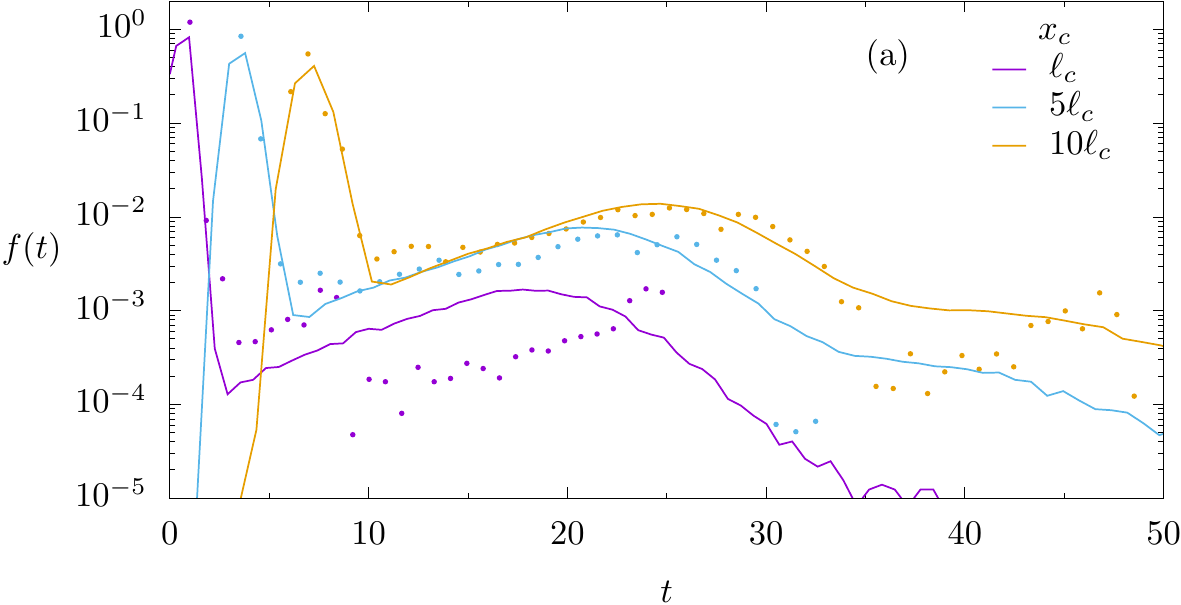}
  \includegraphics[width=.45\textwidth]{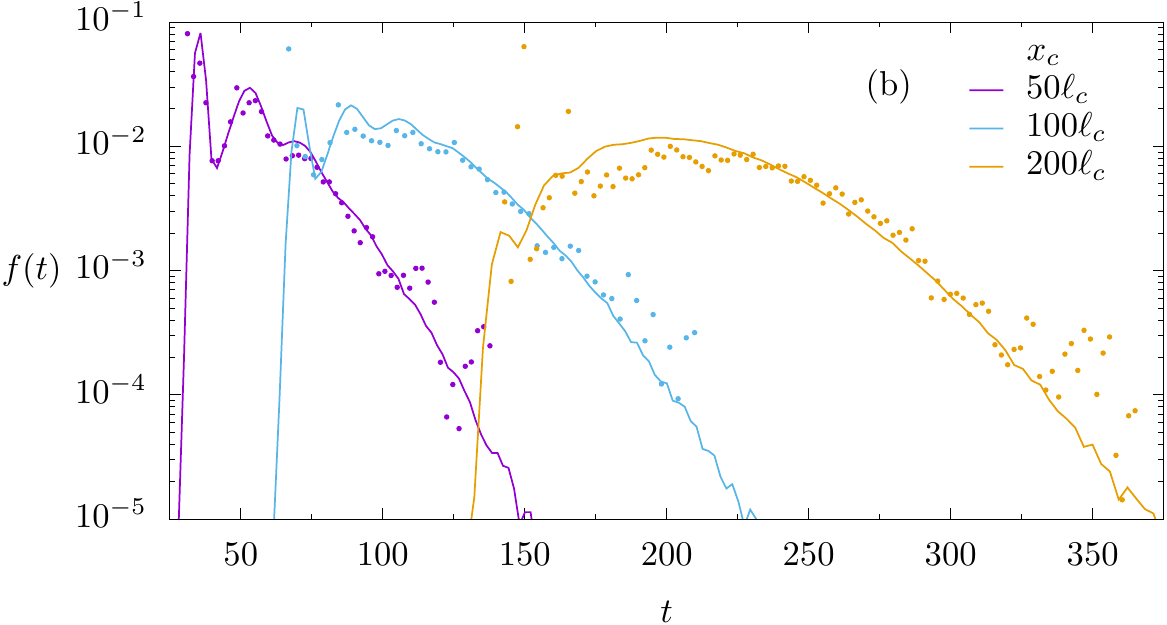}
  \includegraphics[width=.45\textwidth]{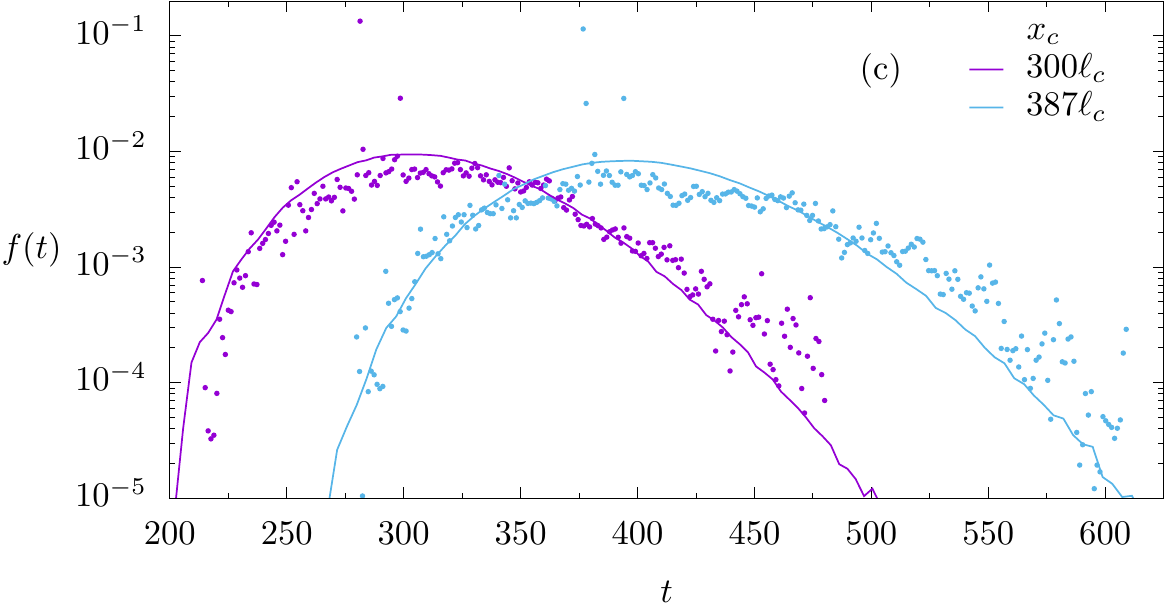}
  \includegraphics[width=.45\textwidth]{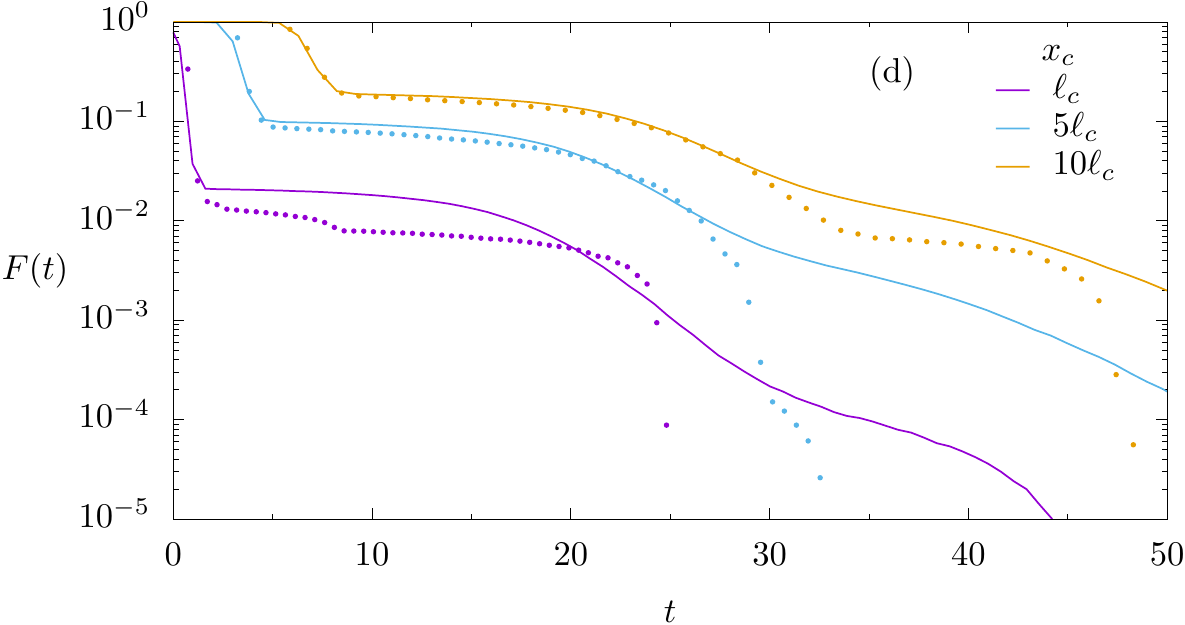}
  \includegraphics[width=.45\textwidth]{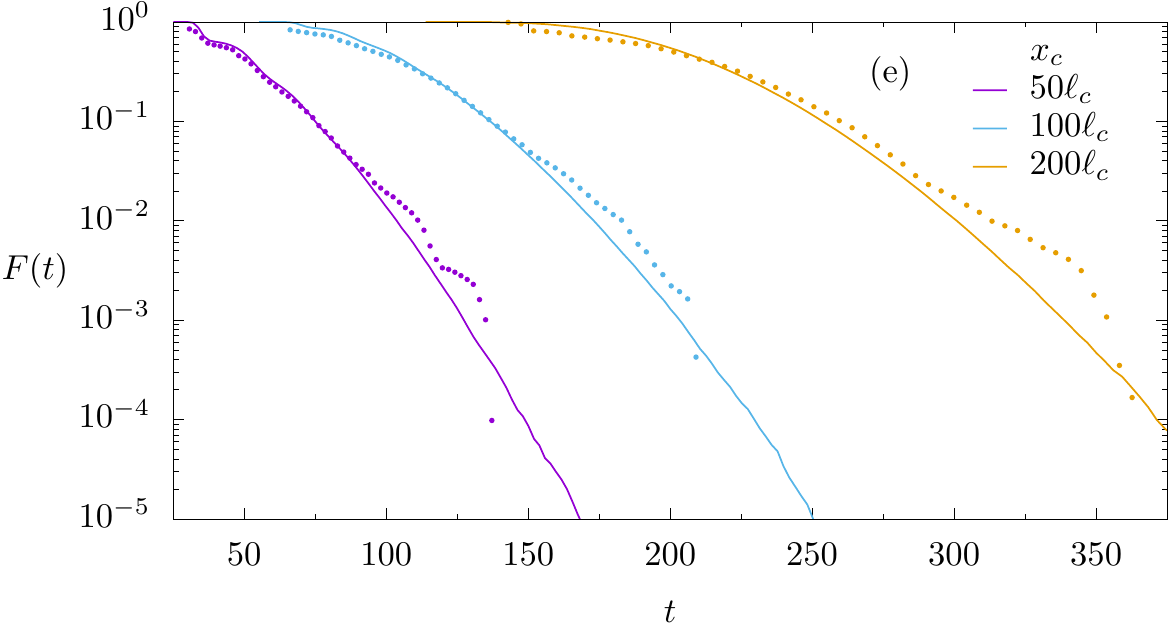}
  \includegraphics[width=.45\textwidth]{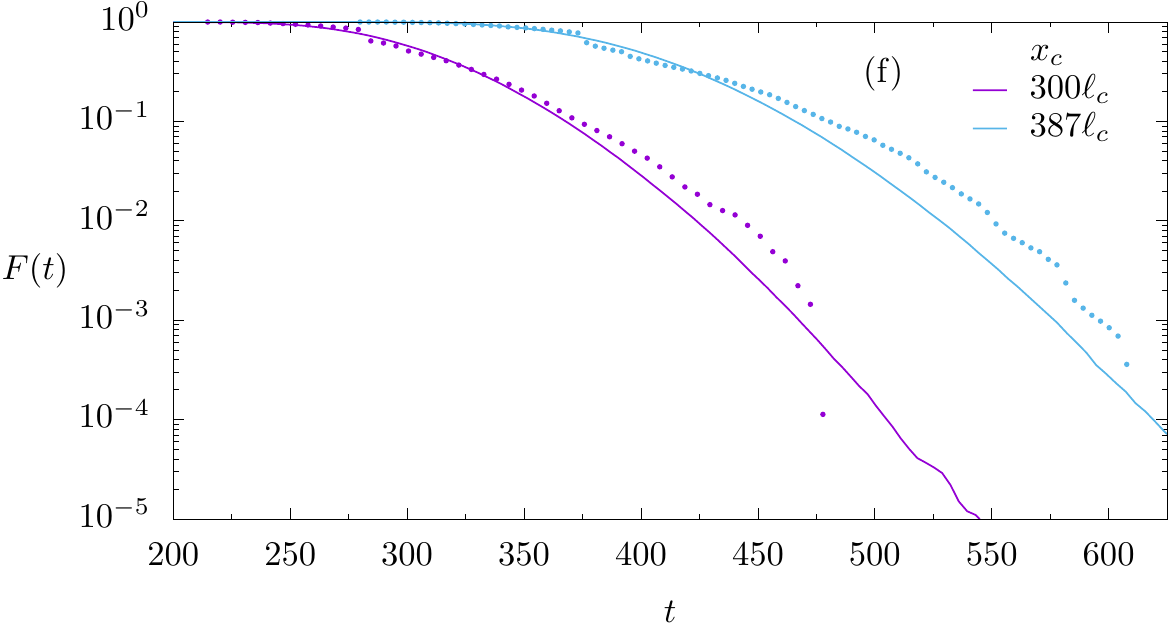}
  \caption{Comparison of the breakthrough curves (a--c) and
    complementary cumulative breakthrough curves (d--f) of the CTRW
    model \eqref{eq:ctrw} (solid lines) to the numerical simulations (dots) for a
    random arrangement of inclusions with $L_{x}=387 \ell_c$,
    $L_{y}=3.8 \ell_c$, $\kappa = 0.01$, and $\chi=0.3$. The curves are measures at
    increasing distance from the inlet. The CTRW models uses the
    parameters (velocity in the matrix, mean number of trapping
    events) measured at the outlet. The mean number of trapping events
    is rescaled for the intermediate distances.}\label{fig:CTRW-5100}
\end{figure}

Before we apply this model to the data of the direct numerical
simulations, some comments on its assumptions are in order. The basis
of the model is the assumption of ergodicity of the underlying disorder
in the following sense. First, the CTRW samples the number $n_{s}$ of
trapping events from a Poisson distribution. This implies that the
inclusion pattern is random and fluctuates on a characteristic length
scale. All particles sample from the same Poisson distribution, this
means all particles must have access to the same statistics as they
move through the medium, which means that the spatial pattern needs to
be stationary. The same holds for the distribution of trapping times,
which are sampled as independent identically distributed random
variables. In the following, we analyze the breakthrough curves in the
light of these remarks.

Figure \ref{fig:CTRW-5100} compares the results for the breakthrough
curves of the direct numerical simulations to the prediction of the
CTRW for a narrow medium of $L_y = 3.8 \ell_{c}$ at different
distances from the inlet. This medium has in average 2.5 inclusions
per vertical cross section. For this case, we do not expect that the
upscaled model provides a good prediction at short distances because
the ergodicity conditions discussed above do not apply. All the
particles in the direct numerical simulation initially experience the
same or similar disorder, this means they are not independent
statistically. In fact, transport can be interpreted as occurring in
streamtubes as discussed above. Only with distance from the inlet,
particles start sampling the medium structure and heterogeneity. This
means in terms of the number of times particles pass an inclusion, and
the trapping times experienced. Remarkably, sampling is sufficiently
efficient due to the random nature of the medium that the upscaled
CTRW model reproduces the primary peak of first arrival and secondary
peaks that correspond to different numbers of trapping events. This
means, particles sample along single streamlines a representative part
of the medium statistics.
\begin{figure}
  \centering
  \includegraphics[width=.49\textwidth]{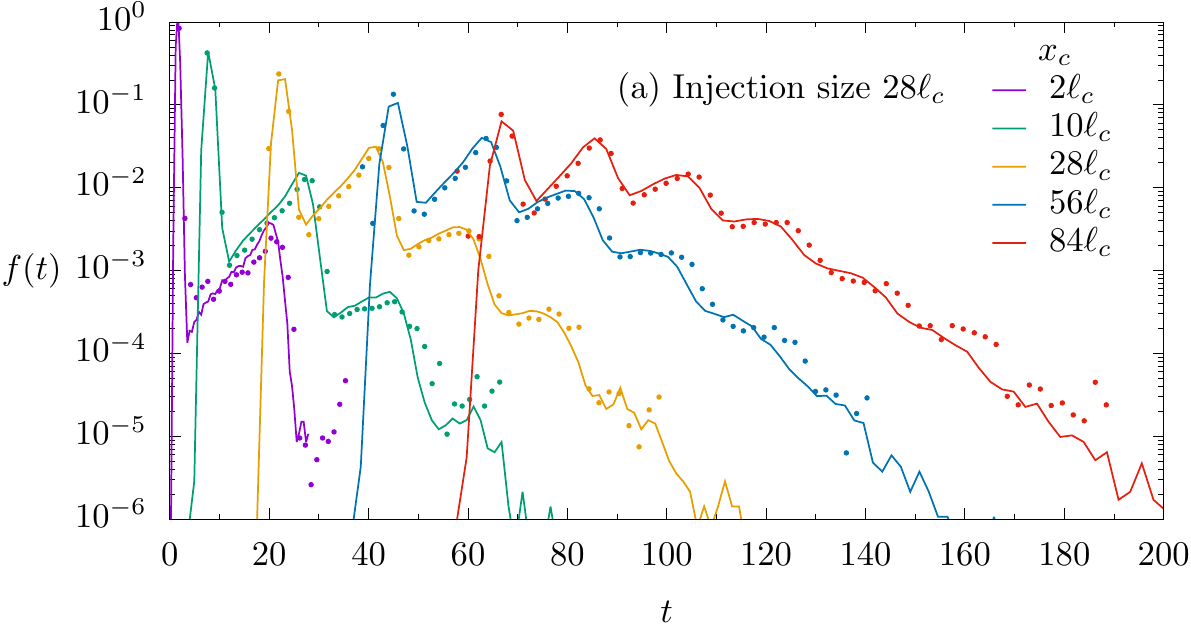}
  \includegraphics[width=.49\textwidth]{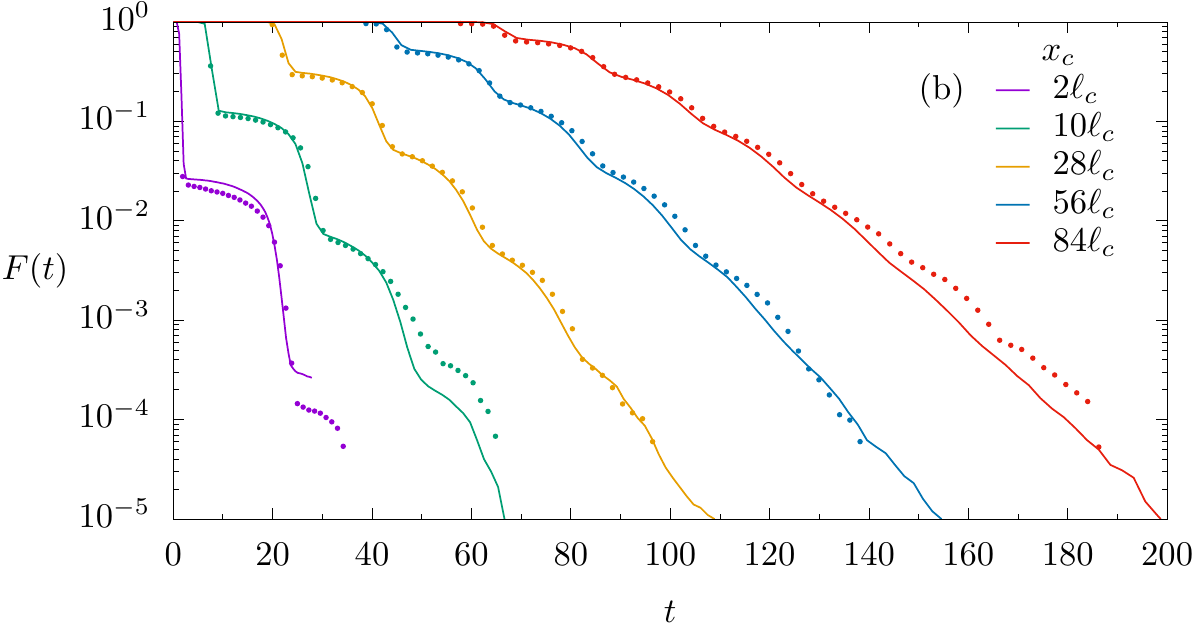}
  \caption{Comparison of the breakthrough curves (a) and complementary
    cumulative breakthrough curves (b) of the CTRW model
    \eqref{eq:ctrw} (solid lines) to the numerical simulations (dots)
    for a random arrangement of inclusions with $L_{x} = 84\ell_{c}$,
    $L_{y} = 28\ell_{c}$, $\kappa = 0.01$, and $\chi=0.2$ for an
    injection line of size equal to the domain height. The curves are
    measured at different distances from the
    inlet.}\label{fig:CTRW-Ly}
\end{figure}

The breakthrough curves shown in Figure \ref{fig:CTRW-Ly} are measured in a
medium whose lateral extension comprises $28 \ell_c$, this means, the
particles injected over the medium cross section sample from the start a
representative part of the medium heterogeneity, both in terms of the
spatial structure and in terms of the trapping time statistics. Thus,
the upscaled CTRW model predicts the direct simulation data already
at short distances from the inlet. It provides good predictions for
the first arrival, and also, as in Figure~\ref{fig:CTRW-5100} for the
secondary peaks.
\section{Conclusions}
We studied in this paper the advective transport of solutes in an
idealized heterogeneous porous medium consisting of a homogeneous
background material with low permeability circular inclusions. In such
media the distribution of flow, and solute if injected uniformly,
between the matrix and the low permeability inclusions is given by the
permeability ratio provided that the inclusion density is not very
high. While velocity in the matrix depends on fraction of the area
covered by the inclusions, the mean velocity in the inclusions follows
a lognormal distribution.

Transport is characterized by breakthrough curves whose shape evolves
as the medium is sampled. At short distance from the injection inlet,
or when the injection length is smaller than the domain, the
breakthrough curve has a wavy shape that reflects the trapping of
particles at the inclusions. This curve can be interpreted as
transport through streamtubes with different velocities. As the
distance, or injection length, grows, the properties of the medium are
better sampled and the curve becomes smoother. Particles arrive
gradually at the control plane, which reflects the tortuosity of the
streamlines that go through the matrix. The better sampling of the
velocity distribution in the inclusions makes the tail of the curve
also smoother. These features are common to the behavior of the ADE
\eqref{eq:ADE}. However, the shape of the curves cannot be predicted
with a macrodispersion coefficient. The ADE overestimates
concentration at early times and underestimates it at late
times. Unrealistic values of the dispersion coefficient are obtained
from the variance of the breakthrough times (see
Table~\ref{tab:velDisp}). This is particularly accentuated when the
medium is undersampled.

The problem with the representation of variable travel times as
macrodispersion can be illustrated with the results of
\citet{Eames1999}. In their analysis, the macrodispersion coefficient
derived for a finite permeability ratio $\kappa$ diverges in the
limit $\kappa \to 0$. However, if the inclusions are impermeable from
the beginning, the macrodispersion coefficient is finite. The latter
captures the effect of tortuous streamlines in the background
material. If inclusions are considered permeable and $\kappa$ is very
high, the macrodispersion coefficient also captures the trapping in
the inclusions.  As long as the inclusions are permeable, there is a
probability that solute gets into an inclusion. If inside the
inclusion, solute is slowed down, which leads to a tailing of the
breakthrough curve. The lower the permeability ratio, the stronger
the tailing due to the longer delay time. Therefore, in the limit of a
ratio of zero, the tail gets infinitely long. The effect that the
probability to get into an inclusion also goes to zero does not
counteract the infinitely long trapping time. As the macrodispersion
coefficient is obtained from the total solute distribution in the
domain, a retention in an inclusion for infinite time leads to a
diverging macrodispersion coefficient. In case that the inclusions are
impermeable from the beginning, there is no transport through
inclusions that could cause tailing. Therefore the macrodispersion
coefficient is finite. The behavior is inherent to assuming an
advection dispersion equation for the upscaled model.

We developed an upscaled transport model using a Lagrangian
framework. The main assumption of the model is that the medium structure is
ergodic. Therefore the model performance improves, as the
particles sample a larger part of the medium heterogeneity. This is
the case either when the particles travel a long distance from the inlet or when the injection
length is long, so that the medium properties are explored even at short
distance from the inlet.

The CTRW model developed has solid predictability capabilities because
it is parameterized by measurable medium properties. The CTRW model is
parameterized by the trapping rate, which we observed can be
characterized by a Poisson distribution whose trapping rate (number of
trapping events per distance) is inversely proportional to the mean
separation between inclusions, the velocity in the matrix, which is
well approximated by a function of the area covered by the inclusions
(Figure~\ref{fig:vel-mat}), and the velocity distribution inside the
inclusions. The mean velocity inside the inclusions follows a
log-normal distribution, which needs further investigation. So does
the Poisson distribution of trapping events, which constitutes an
important part of the upscaled model with possible applications to
more general scenarios. Furthermore, we assumed constant porosity and considered a
2D scenario. We anticipate that for 3D geometries and variable porosity, the trapping rate
and velocity distribution may change, and that otherwise the derived model
remains valid.

The upscaled model was also formulated in an equivalent
mobile-immobile memory function
model. The memory function is determined by the trapping time
distribution and for the same reasons as outlined above for the CTRW
model, predictable from information about parameters and structure of
the porous medium. The memory function in our setting describes
tailing of breakthrough curves due to advective transport through
circular inclusions with low permeability. To generalize it towards
media with inclusions with a distribution of permeability values or
sizes, is straight forward, if appropriate models for the velocity
distribution inside of the inclusions can be formulated.

Purely advective transport was here considered as a limiting case for
advective-diffusive transport. The other limiting case, purely
diffusive transport inside of inclusions, has been studied and mobile-immobile
models are well established for it. In a next step it would be
necessary to consider the combined effect of advective and diffusive
transport inside of inclusions, and to derive predictive
mobile-immobile memory function models based on the models for pure advection and for pure
diffusion.
%
%
\section*{Acknowledgements}
Data used for producing the figures can be downloaded from
digital.csic.es (https://digital.csic.es/handle/10261/216991) and by
solving the respective equations. The authors thank Prof. Timothy
R. Ginn and two anonymous reviewers for their comments on the
paper. J.J.H. and M.D. acknowledge the support of the Spanish Ministry
of Science and Innovation (project CEX2018-000794-S and project
HydroPore PID2019-106887GB-C31). J.J.H. acknowledges the support of
the European Social Fund and Spanish Ministry of Ministry of Science,
Innovation and Universities through the ``Ram\'on y Cajal'' fellowship
(RYC-2017-22300). The authors thank Pierre Uszes for providing the
code to measure the average distance between inclusions.
%
%
\section*{Declaration of Interests}
The authors report no conflict of interest.
\appendix
%
%
\section{Upscaling\label{app:upscaling}}
We note that the number $n_L$ of inclusions within a distance $L$ between
inlet and outlet is a Poissonian random variable. We set the average
time spent mobile equal to
\begin{align}
s = \frac{L}{v_{m}}.
\end{align}
Thus we can set $L = v_{m} s$ and $n_s \equiv n_{L}$. Accordingly, we set
the immobile time $\tau_{im}(s)$ after the mobile time $s$ has passed equal to
\begin{align}
\tau_{im}(s) = \sum_{i = 1}^{n_{s}} \frac{\ell_{i}}{v_{i}} - \frac{\ell_{i}}{v_{m}}.
\end{align}
where the distance $\ell_{i}$ traveled across an inclusion are distributed
according to~\eqref{eq:p_li}. The inclusion velocity $v_{in}$ is
lognormally distributed, see Section~\ref{sec:vpdf}.
The second term under the sum compensates for the fact that the
average mobile time accounts for the full distance $L$ and not only
for the distance $L - \sum_{i = 1}^{n_{s}} \ell_{i}$ a particle moves in
the matrix. With this reasoning, we obtain expression~\eqref{eq:time} for
the clock time $t(s)$.

Next we consider expression~\eqref{eq:psi} for the Laplace transform of
the distribution $\psi(t|s)$ of clock time $t(s)$. It can be written
as
\begin{align}
  \psi(t|s) = \left\langle \delta\left(t - s -  \sum\limits_{i = 1}^{n_s} \tau_{i}\right) \right\rangle,
\end{align}
which we can expand as
\begin{align}
\label{app:psi}
  \psi(t|s) = \sum_{n = 0}^\infty \left\langle \delta\left(t - s
  - \sum\limits_{i = 1}^{n} \tau_{i}\right) \delta_{n,n_s} \right\rangle
  = \sum_{n = 0}^\infty \left\langle \delta\left(t - s -
  \sum\limits_{i = 1}^{n} \tau_{i}\right)\right\rangle p_{n}(s),
\end{align}
where $p_{n}(s)$ is the Poisson distribution
\begin{align}
\label{app:poisson}
p_{n}(s) = \frac{(k v_m s)^n\exp(-k v_m s)}{n!}.
\end{align}
The Laplace transform of~\eqref{app:psi} is
\begin{align}
\psi^{\ast}(\lambda|s) = \exp\left(-\lambda s \right) \sum_{n = 0}^\infty \left\langle \exp\left(-\lambda \sum\limits_{i = 1}^{n} \tau_{i}\right)\right\rangle p_{n}(s),
\end{align}
which can be written as
\begin{align}
\psi^{\ast}(\lambda|\ell_{c}) = \exp\left(-\lambda s \right) \sum_{n = 0}^\infty \psi_{f}^\ast(\lambda)^n p_{n}(s)
\end{align}
because the $\tau_{i}$ are independent identically distributed random numbers.
Inserting now expression~\eqref{app:poisson} for the Poisson distribution gives
\begin{align}
  \psi^{\ast}(\lambda|s) = \exp\left(-\lambda s \right) \sum_{n =
  0}^\infty \psi_{f}^\ast(\lambda)^n \frac{(k v_m s)^n \exp(-k v_m s)}{n!}
\end{align}
The exponential sum can be evaluated explicitly and thus
\begin{align}
\label{app:psi}
  \psi^{\ast}(\lambda|s) = \exp\left(- \lambda s - k v_m s [1 - \psi^{\ast}_{f}(\lambda)]\right).
\end{align}

The concentration distribution in the CTRW framework can be written as
\begin{align}
\label{eq:ctrw1}
c(x,t) = \langle \delta[t - t(s)] \rangle = \int\limits_0^\infty ds
  c_0(x,s) h(s,t),
\end{align}
where we defined
\begin{align}
c_0(x,s)\langle = \delta[x - x(s)] \rangle, && h(x,s) = \langle \delta[s - s(t)] \rangle
\end{align}
The distribution $c_0(x,s)$ satisfies the advection-dispersion equation
\begin{align}
\label{app:ade}
\frac{\partial c_0(x,s)}{\partial s} + v_m \frac{\partial
  c_0(x,s)}{\partial x} - D_m \frac{\partial^2 c_0(x,s)}{\partial x^2}
  = 0.
\end{align}
The distribution $h(s,t)$ of the renewal process $s(t) = \max(s|t(s)
\leq t)$ satisfies
\begin{align}
\int\limits_0^s ds h(s,t) = \int\limits_t^\infty dt' \psi(t'|s),
\end{align}
Thus, the concentration $c(x,t)$ can be written as
\begin{align}
\label{eq:ctrw2}
c(x,t) = \int\limits_0^\infty ds c_0(x,s) \frac{\partial}{\partial s} \int\limits_t^\infty dt' \psi(t'|s),
\end{align}
This equation can be written in Laplace space as
\begin{align}
\label{eq:ctrw3}
c^\ast(x,\lambda) = \int\limits_0^\infty ds c_0(x,s)
  \frac{\partial}{\partial s} \frac{1 -
  \psi^{\ast}(\lambda|s)}{\lambda}.
\end{align}
Inserting expression~\eqref{app:psi} for $\psi^{\ast}(\lambda|s)$ gives
\begin{align}
\label{eq:ctrw3-2}
c^\ast(x,\lambda) = \left\{\lambda + k v_m [1 -
  \psi^{\ast}_{f}(\lambda)] \right\} \int\limits_0^\infty ds c_0(x,s)
  \psi^{\ast}(\lambda|s).
\end{align}
On the other hand, integration of~\eqref{eq:ctrw3} by parts gives
\begin{align}
\label{eq:ctrw4}
\lambda c^\ast(x,\lambda) = - \int\limits_0^\infty ds \frac{\partial
  c_0(x,s)}{\partial s} [1 -
  \psi^{\ast}(\lambda|s)] =  \delta(x) + \int\limits_0^\infty ds \frac{\partial
  c_0(x,s)}{\partial s} \psi^{\ast}(\lambda|s)
\end{align}
where $c_0(x,s=0) = \delta(x)$. Equation~\eqref{app:ade} implies that
the right side can be written as
\begin{align}
\label{eq:ctrw5}
\lambda c^\ast(x,\lambda) =  \delta(x) +
  \left[ + v_m \frac{\partial}{\partial x} - D_m
  \frac{\partial^2}{\partial x^2} \right] \int\limits_0^\infty ds c_0(x,s) \psi^{\ast}(\lambda|s)
\end{align}
Using now expression~\eqref{eq:ctrw3-2} on the right side of this
expression in order to eliminate $c_0(x,s)$ in favor of
$c^\ast(x,\lambda)$ gives
\begin{align}
\label{eq:ctrw6}
\lambda c^\ast(x,\lambda) =  \delta(x) +
  \left[ + v_m \frac{\partial}{\partial x} - D_m
  \frac{\partial^2}{\partial x^2} \right] \frac{c^\ast(x,\lambda)}{\lambda - k v_m [1 -
  \psi^{\ast}_{f}(\lambda)]}
\end{align}
We define now the mobile concentration $c_m^\ast(x,\lambda)$ as
\begin{align}
c_m^\ast(x,\lambda) = \frac{c^\ast(x,\lambda)}{\lambda + k v_m [1 -
  \psi^{\ast}_{f}(\lambda)]}.
\end{align}
Thus, we obtain for the mobile concentration $c_m^\ast(x,s)$ the
governing equation
\begin{align}
\lambda c_m^\ast(x,\lambda) + \lambda \varphi^\ast(\lambda) c_m^\ast(x,\lambda)
=  \delta(x) + \left[ + v_m \frac{\partial}{\partial x} - D_m
  \frac{\partial^2}{\partial x^2} \right] c_m^\ast(x,\lambda),
\end{align}
where we defined the memory function
\begin{align}
\varphi^\ast(\lambda) = k v_m \frac{1 -
  \psi^{\ast}_{f}(\lambda)}{\lambda}.
\end{align}
Furthermore, we can now define the immobile concentration $c_{im}^\ast(x,\lambda)$ as
\begin{align}
c_{im}^\ast(x,\lambda) = k v_m [1 - \psi^{\ast}_{f}(\lambda)]
  c_m^\ast(x,\lambda).
\end{align}
The inverse Laplace transform of this expression is given
by~\eqref{eq:phi}.
%
%
\bibliographystyle{jfm}
\bibliography{Manuscript}
%
%
\end{document}